\documentclass[reprint,amsmath,amssymb,aps]{revtex4-2}

\usepackage{graphicx}
\usepackage{dcolumn}
\usepackage{bm}

\usepackage{xcolor}
\usepackage{cancel}
\usepackage{tocloft}

\usepackage{hyperref}

\begin{document}

\preprint{APS/123-QED}

\title{Phonon controlled mechanical memory\\via pinning and depinning of transition waves
}

\author{Samuele Ferracin$^1$}
\author{Dengge Jin$^1$}%
\author{Vincent Tournat$^{2,3}$}
\email{email: vincent.tournat@univ-lemans.fr}
\author{Jordan R. Raney$^1$}%
\email{email: raney@seas.upenn.edu}
\affiliation{$^1$Department of Mechanical Engineering and Applied Mechanics, University of Pennsylvania, Philadelphia, PA 19104}
\affiliation{$^2$Laboratoire d'Acoustique de l'Université du Mans (LAUM), Institute d'Acoustique – Graduate School, CNRS, Le Mans Université, France}
\affiliation{$^3$John A. Paulson School of Engineering and Applied Sciences, Harvard University, Cambridge, MA 02138, USA}


\begin{abstract}
Multistable mechanical metamaterials enable programmable transitions between discrete stable states through propagating kink transition waves (TWs). Yet controlling these kinks typically requires local actuation or high-energy deformation, limiting scalability. Here we demonstrate a universal strategy for pinning and depinning TWs using local defects and boundary phonon excitations. Inspired by phonon–dislocation interactions in crystalline solids, we use pairs of phonons that form a beating envelope resonant with the pinned kink’s translational mode, which lies within a phononic band gap. This resonant coupling efficiently transfers energy to the kink, allowing it to overcome defect barriers and propagate across impurities. The proposed mechanism enables application of these systems as information processing units in mechanical computing, namely as 
scalable and more robust mechanical memory.

\end{abstract}

\maketitle




Multistable mechanical metamaterials (MMMs) have emerged as a powerful platform for programmable matter, capable of discrete local or global state transitions, enabling shape morphing \cite{dudek2025shape}, tunable stiffness \cite{tahidul2024reprogrammable}, impact mitigation \cite{liang2025ideal}, and wave control \cite{nadkarni2016unidirectional}. These metamaterials have been used in mechanical computing~\cite{song2019,yasuda2021,he2024programmable,chen2025}, and can substitute, or interact with, conventional computing architectures to create intelligent systems~\cite{jiang2019, liu2023cellular}. MMMs 
can sense and compute responses based on pre-programmed logic \cite{mei2021mechanical}, and can pre-process physical information to alleviate the workload on electronic systems \cite{byun2024integrated}.
Mechanical memory units are necessary components of these systems, enabling the metamaterial’s behavior to depend on loading history by storing information as states of local deformation \cite{yasuda2017origami, jules2022delicate, stenseng2025bi, bodaghi20194d}.

In MMMs, transitions between these states are often mediated by propagating fronts known as kink (and anti-kink) transition waves (TWs)~\cite{raney2016stable, puglisi2000mechanics, abeyaratne2006evolution, nadkarni2016unidirectional}.
The individual multistable units can possess symmetric or asymmetric on-site potentials. 
In the former case, TW propagation distance is limited by dissipation and finite input energy, frequently resulting in incomplete transitions \cite{peyrard1984kink}. Static kinks in such systems can also be insufficiently stable to resist unwanted perturbations in practical applications~\cite{hasenfratz1977interaction}, limiting the utility of symmetric potentials as mechanical memory. 
In contrast, lattices with asymmetric bistable potentials can support self-sustained propagation even in dissipative environments: the potential asymmetry releases stored energy, enabling robust propagation and full reconfiguration~\cite{nadkarni2016unidirectional,raney2016stable}. Yet, such systems can access only a limited number of global configurations, determined by the number of stable equilibria in the on-site potential. 
To overcome this limitation and engineer reprogrammable metamaterials with enhanced reconfigurability, one can introduce local defects that arrest (pin) TW propagation at prescribed locations \cite{deng2020characterization, wang2023phase2} or change the propagation path \cite{jin2020guided}. These defects act as energy barriers that define programmable global states. While pinned kinks can be more stable than the symmetric case with no impurities, such barriers are typically ``arbitrarily'' stiff, hindering additional reconfiguration. Achieving further reprogramming requires, instead, carefully designed defects that can allow both pinning and controlled depinning of the kink, enabling kink propagation past the impurity.

One limitation of existing strategies for mechanical memory is the need for complex networks of actuators to write new states, or peripheral equipment to sense them \cite{chen2021reprogrammable}. 
Controlling kinks location/memory states through boundary excitations is crucial for scalable reprogrammability, avoiding exponential increases of complexity, actuator networks or embedded stimuli. 
Prior approaches demonstrated the use of simple boundary excitations using periodic quasi static compression/tension \cite{pechac2023mechanical}, but they require specific metamaterial designs coupled to stiff substrates and high energy deformations.
Another approach demonstrated that arbitrary memory states can be encoded remotely by sending tailored nonlinear waves from the system's edge to switch specific internal bits without local actuation \cite{watkins2025arbitrary}. However, this approach requires a precise waveform, amplitude, and frequency, as the system's complex landscape makes it susceptible to any small deviations in the input signal. 

In this work, we propose a boundary excitation inspired by processes in crystalline solids, whereby phonons can promote or suppress dislocation mobility, altering plasticity and thermal conductivity \cite{currie1979dynamics, cuevas2014sine, swinburne2013theory, chen2017effects, shilo2003stroboscopic}, and external ultrasonic excitations can enhance mobility via acoustoplasticity (Blaha effect) \cite{siu2011understanding}. 
We leverage the ability of phonons to couple with kink solitons and control their position \cite{hasenfratz1977interaction} in a mechanical metamaterial, extending previous works to systems with the dissipation, impurities, or asymmetries that are needed for reliable mechanical memories.
We also propose a new phonon-mediated depinning mechanism using pairs of phonons that generate a beating envelope resonant with the pinned kink's translational (T) mode, which resides in a band gap. This nonlinear beating efficiently transfers energy into kink motion, overcoming defect barriers and enabling progressive reconfiguration, greatly expanding the accessible configuration space of multistable lattices. This approach could be used as a universal strategy for deterministically writing new memory states by pinning and depinning kinks across a broad class of bistable metamaterials. The strategy results in particularly robust memory because depinning can be achieved only with specific forcing modulations, as will be discussed below, so external noise and perturbations cannot randomly overwrite memory states.

Below, we first 
outline the transition wave dynamics in an unperturbed lattice with no impurities. We then determine the minimum defect strength to pin the kink for various damping regimes. Finally, we analyze the vibrational modes of the pinned kink and the coupling between phonons and kinks to develop a nonlinear reduced-order model (ROM) capturing the dynamics of the oscillating kink and resonant de-pinning.

\begin{figure}[b]
\includegraphics[width=\columnwidth]{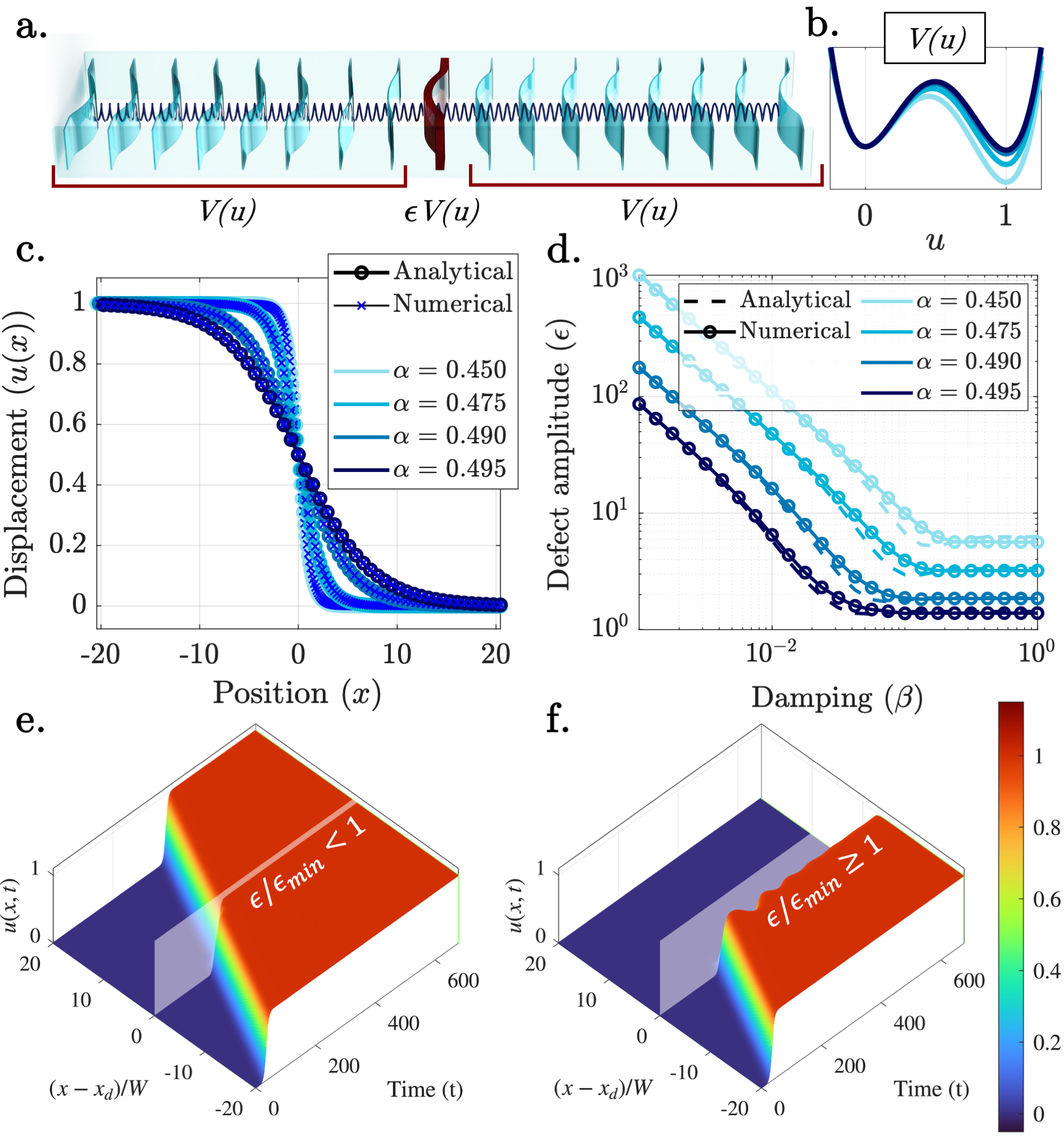}
\caption{\label{fig:fig1} (a) Possible mechanical realization of the system, with bistable beams coupled by springs, with defect at the center. (b) Symmetric and asymmetric potentials $V(u)$. (c) Transition wave displacement profile, for different asymmetries $\alpha$.
(d) Threshold value $\epsilon_{min}$ for various damping coefficients, at different $\alpha$.
(e) TW passes through a defect which $\epsilon/\epsilon_{min}<1$, and (f) is pinned if $\epsilon/\epsilon_{min}\geq1$. }
\end{figure}

We consider a 1D continuous periodic lattice comprising bistable units with linear nearest-neighbor interaction (Fig.~\ref{fig:fig1}a). We select this system because it can be used as a base approximation for multiple different and more complex metamaterial architectures supporting TWs \cite{yasuda2023nucleation, deng2022topological, deng2021nonlinear,paliovaios2024transition,jiao2024nucleation} . 
The displacement field $u(x,t)$ evolves according to
\begin{equation}\label{eq: EOMtot}
\ddot{u} + \beta \dot{u} - c^2 u_{,xx} + \Big(1 + (\epsilon - 1)\delta(x - x_d)\Big)\frac{dV}{du} = 0,
\end{equation}
where $\beta$ is a viscous damping parameter, $c$ is the coupling stiffness, and the bistable on-site potential $V(u)$ (with minima at $u=0,1$) is described as:
\begin{equation}
    V(u) = \frac{1}{4} u^4 - \frac{\alpha + 1}{3} u^3 + \frac{\alpha}{2} u^2 .
\end{equation}
Here $0\leq\alpha\leq1$ determines the potential's asymmetry ($\alpha=0.5$ represents a symmetric energy landscape, see Fig. \ref{fig:fig1}b).
We define an impurity at location $x_d$ by locally modifying the on-site potential. $\delta$ is a regularized Dirac delta and $\epsilon>1$ defines the defect strength. This modification produces an effective repulsive interaction potential that slows the approaching kink, and may stop its propagation if sufficiently strong. Other types of defects could be considered, e.g., a mass defect or a soft defect ($\epsilon<1$), but these have some disadvantages for our purposes. A mass defect could transiently slow the kink down, but without static pinning. Soft defects promote resonant transfer of energy between kink and localized defect modes that could trap kinks, but the behavior is complex and sensitive to initial conditions (e.g., kink speed)~\cite{malomed1993interactions, fei1992resonant}, making it difficult to reliably define a threshold defect strength.

\begin{figure}[b]
\includegraphics[width=\columnwidth]{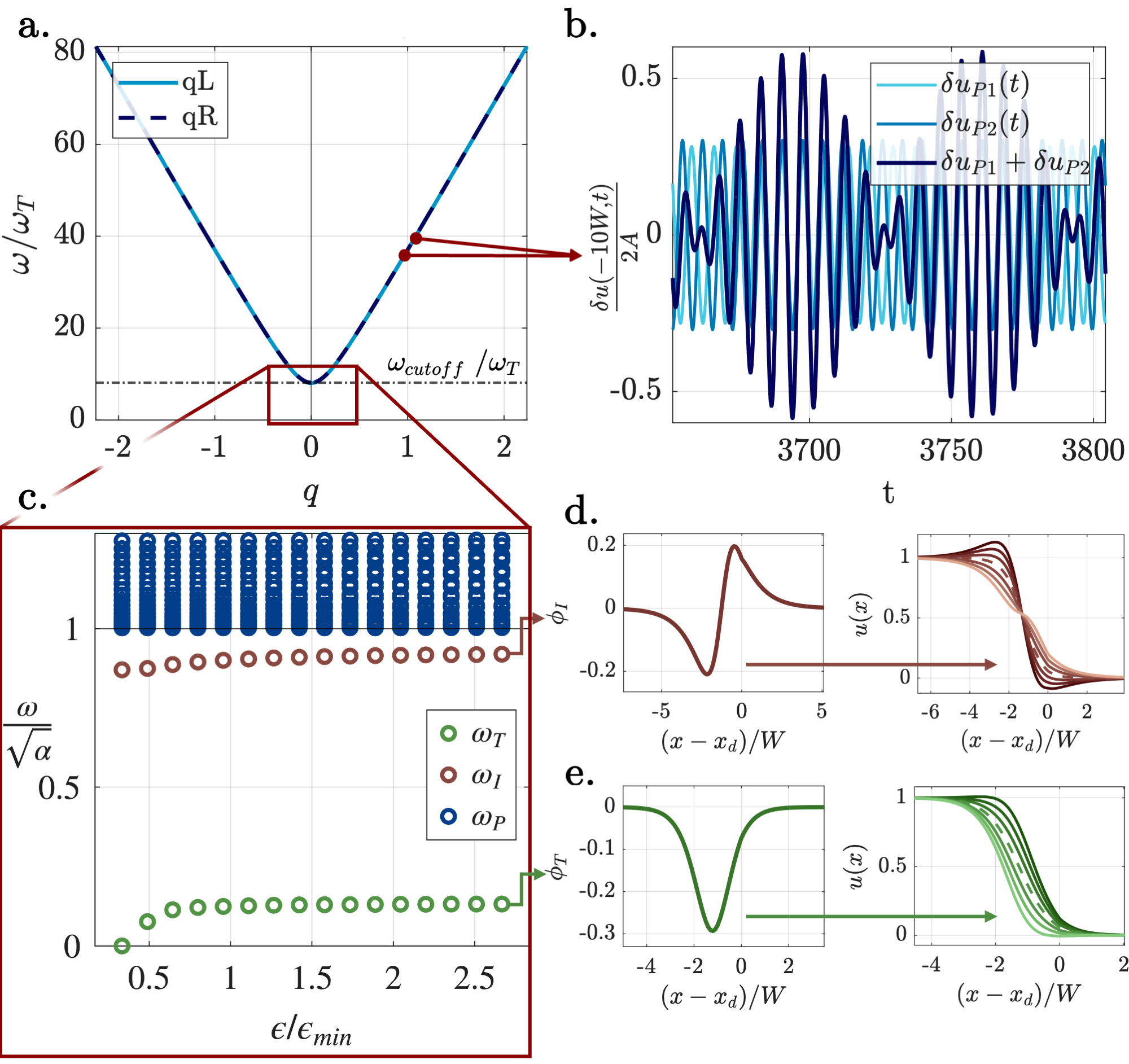}
\caption{\label{fig:fig2} (a) Dispersion relation of the system. (b) Beating perturbation as seen from a unit far before the pinned kink. The two separate phonons are indicated in light blue and the beating that they generate in dark blue. (c)~Spectrum of the localized and extended modes of the pinned kink. In green the translational T-mode (low frequency) and in red the internal I-mode, in dark blue the extended modes. $\epsilon<0.3\epsilon_{min}$ defects are not able to statically hold the kink (showing $\omega_T=0$). (d) The I-mode and (e) T-mode shape (left) and the displacement perturbation they cause with different intensities $a_J$ (right); the dashed curves correspond to the unperturbed kink when $a_J=0$. A second-order discontinuity in the shape mode can be seen at $x-x_d=0$ because of the defect.}
\end{figure}

Considering a lattice with no impurities ($\epsilon=1$), we analytically determine the form of the TW (see \cite{wang2023phase} and SI Section~III for the full procedure):
\begin{equation}\label{eq:analytical_kink}
u_k(x,t) = \frac{1}{2} \left[ 1 - \tanh\left( \frac{(x - x_0) - v_0 (t-t_0)}{W} \right) \right] .
\end{equation}
When $\alpha<0.5$, the TW propagates with speed $v_0 = \frac{c (1 - 2\alpha)}{\sqrt{2\beta^2 + (1 - 2\alpha)^2}}$ and characteristic width $W = 2\sqrt{2(c^2-v_0^2)}$, governed by the balance between the energy released during state switching and the system's dissipation. Fig. \ref{fig:fig1}c shows the displacement profile of TWs for various $\alpha$. 
Importantly, linearization about each equilibrium state ($u=0,1$) gives two Klein-Gordon dispersion relations, which present a low frequency band gap, respectively at $\omega<\sqrt{\alpha}$ and $\omega<\sqrt{1-\alpha}$ 
(see Fig.~\ref{fig:fig2}a and Eq. \ref{eq:linearwaves}).

In this work we seek to understand pinning and depinning of transition waves due to interaction with localized defects. We need to calculate the minimum (threshold) value $\epsilon=\epsilon_{min}$ that is sufficient to pin the kink while still allowing it to be easily unpinned. To do this for a broad range of parameters, different damping regimes must be considered: in the low-damping regime ($\gamma \ll 1$, where $\gamma = \beta \frac{W}{v_0}$ is a dimensionless parameter quantifying the effective damping over the interaction timescale), the kink interacts with the defect nearly conservatively. The threshold $\epsilon_{min}^{adiab}$ is determined by equating the kink's kinetic energy $E_k=\frac{v_0^2}{6 W}$ (Eq. \ref{eq:SIEk}) to the peak of the interaction potential $U_d(\xi_0)$ between the defect and the kink (Eq. \ref{eq:SIinteractionpotential}).
Setting $E_k = max(U_d(\xi_0))=(\epsilon_{min}^{adiab}-1)\frac{\alpha^3(2-\alpha)}{12}$ yields:
\begin{equation}
\epsilon_{min}^{adiab} = 1 + \frac{2 v_0^2}{W \alpha^3 (2 - \alpha)}.
\end{equation}
In the high-damping regime ($\gamma \gg 1$), the kink instead interacts quasi-statically with the defect. In this case, the threshold $\epsilon_{min}^{force}$ is found by balancing the driving force $F_{driving}=\frac{1-2\alpha }{12}$ given by the lattice asymmetry (Eq.~\ref{eq:SIdrivingforce}) and the maximum opposing force from the defect $F_{barrier}(\xi_0)=\frac{dV(u(\xi_d-\xi_0))}{du}\frac{du(\xi_d-\xi_0)}{d\xi}\frac{1}{W}$ (Eq. \ref{eq:SIdefectbarrierforce}).

When $F_{driving} + (\epsilon_{min}^{force} - 1) \text{\;max}(F_{barrier}(\xi_0))=0$ (balancing magnitudes),
\begin{equation}
\epsilon_{min}^{force} = 1 - \frac{F_{driving}}{\text{max}(F_{barrier}(\xi_0))}.
\end{equation}

These two regimes are blended together using an exponential weighting which is governed $\gamma$:
\begin{equation}
\epsilon_{min} = \epsilon_{min}^{adiab} e^{-\gamma} + \epsilon_{min}^{force} (1 - e^{-\gamma}).
\end{equation}
This approach captures the damping-dependent pinning threshold well (Fig. \ref{fig:fig1}d), improving the accuracy over a pure energy balance~\cite{malomed1992perturbative}.

Since phonon-assisted depinning relies on propagating waves, this mechanism operates effectively under moderate damping: excessive dissipation attenuates phonons before reaching the kink, whereas very low damping increases the defect's pinning threshold, making depinning more difficult. To balance these opposing effects, we select systems with $\gamma \approx 0.5$ (at slightly lower damping from the kinked part of the plot). The results in the rest of this work are obtained using $\alpha=0.495$, $\beta=0.02$, and $\epsilon=2.8$. Similar results can be obtained for a range of $\alpha$ as explained in SI Section~VIII.

When the kink is pinned, the counteracting forces from the lattice and the defect deform it into a static shape $u_0(x)$ (see SI Fig. \ref{fig:SIu0alpha}) satisfying the static limit of Eq.~\ref{eq: EOMtot}. 
We use a perturbation method to analyze the dynamics of the pinned deformed kink. Perturbing around this state ($u^*=u_0(x)+\delta u(x,t)$), linearizing, and assuming solutions of the form $\delta u(x,t)=\phi(x)e^{\lambda t}$, yields eigenmodes $\phi_j(x)$ and  eigenvalues $\lambda_j$ (see SI Section~V).
The resulting modes are of three types: a finite-frequency translational-like mode $\phi_T$, an internal mode $\phi_I$ (oscillation around the center of the kink), and extended modes (which reside in the pass band with frequencies higher than max$(\sqrt{\alpha}, \sqrt{1-\alpha})$). Fig. \ref{fig:fig2}c-e show the (undamped) frequencies of these modes, as well as their shapes and the deformation of the kink they cause.

To study the dynamics of the pinned kink, we focus on the translational-like (T) mode because it encapsulates the collective displacement of the kink as a whole, effectively governing its motion across the defect. 
This is done by decomposing the perturbation as $\delta u(x,t)=\sum_Ja_J(t) \phi_J(x)$, where $J=\{T, I, P1, P2, ...\}$, $a_J(t)$ are the time-dependent amplitudes of each mode, and $Pj$ refer to the traveling phonon modes derived from a scattering analysis (see SI Section~VI). Then $\delta u(x,t)$ is substituted into the full dynamic equation (Eq. \ref{eq:SIeom_preperturbation}) and projected onto the translational eigenmode. This yields the second order ordinary differential equations (ODE) of a damped linear oscillator of stiffness $\omega_T^2$,
\begin{equation} \label{eq:a_j_ode_linear}
    \ddot{a}_{T}(t) + \beta \dot{a}_{T}(t) + \omega_T^2 a_T(t) = F_T(t) ,
\end{equation}
where the quadratic nonlinearity arising from the potential (see Eq. \ref{eq:SIpotentialterms}) is responsible for the coupling force $F_T(t)$ between the phonons and the T-mode (SI Section~7),
\begin{equation} \label{eq:F_T_definition1}
    F_T(t) = - \int_{-\infty}^{\infty} \phi_T(x) \Big(\frac{V_d'''(u_0)}{2} \sum_J\delta u_{PJ}(x,t)^2 \Big) dx ,
\end{equation}
where $V_d'=\Big(1 + (\epsilon - 1)\delta(x - x_d)\Big)\frac{dV}{du}$.

Because this is a linear oscillator of natural frequency $\omega_T$, which is inside the band gap as shown in Fig.~\ref{fig:fig2}c, forcing the pinned kink with phonons in the propagating band ($\omega>max(\alpha,1-\alpha)\gg\omega_T$) is not an effective way of resonantly injecting energy in the T-mode to achieve depinning by overcoming the interaction potential from the defect.
Instead, we nonlinearly excite $\omega_T$ through the interaction of two propagating phonons at frequency $\omega_1$ and $\omega_2$ to 
create a beating frequency $\Delta \omega = |\omega_1-\omega_2|=\omega_T$ (Fig.~\ref{fig:fig2}b).

Introducing the beating perturbation $\delta u_B(x,t) = \delta u_{P1}(x,t) + \delta u_{P2}(x,t)$ in the forcing term of Eq.~\ref{eq:F_T_definition1}, and separating the terms with different frequencies arising from the quadratic nonlinearity in $F_T$, the total force can be divided into a static, resonant, and non-resonant component:
\begin{equation} \label{eq:aP_ode_forced}
\begin{split}
        F_T(t) =F_{Tstat}+F_{Tres}(t)+F_{Tnon-res}(t)
        \\
        \approx \Big[  a_{P1}^2 K_{\text{stat}}^{11} + a_{P2}^2 K_{\text{stat}}^{22}\Big] +
        \\
        + a_{P1} a_{P2} \Big[ K_{\text{cos},-}^{12} \cos(\Delta \omega t) + K_{\text{sin},-}^{12} \sin(\Delta \omega t) \Big] ,
\end{split}
\end{equation}
where $a_{P1}$ and $a_{P2}$ are the amplitudes of the two phonons. The high-frequency ($F_{Tnon-res}(t)$) is not further considered because it is far from $\omega_T$.
To achieve the maximum response amplitude, and achieve depinning with the lowest possible excitation energy, the beating phonons are selected to maximize the coupling coefficient $(K_{\text{cos},-}^{12})^2 + (K_{\text{sin},-}^{12})^2$.
Significantly, the static component is always positive, so the oscillating force $F_T(t)$ has a positive mean (as shown in Fig.~\ref{fig:SIFt}); physically, this corresponds to a bias that pulls the kink toward the excitation source more strongly than it pushes it forward. While such a bias is unfavorable for depinning the kink from a defect, it can be beneficial for slowing down the propagating TW and facilitate pinning, as will be explained later.

\begin{figure}[b]
\includegraphics[width=\columnwidth]{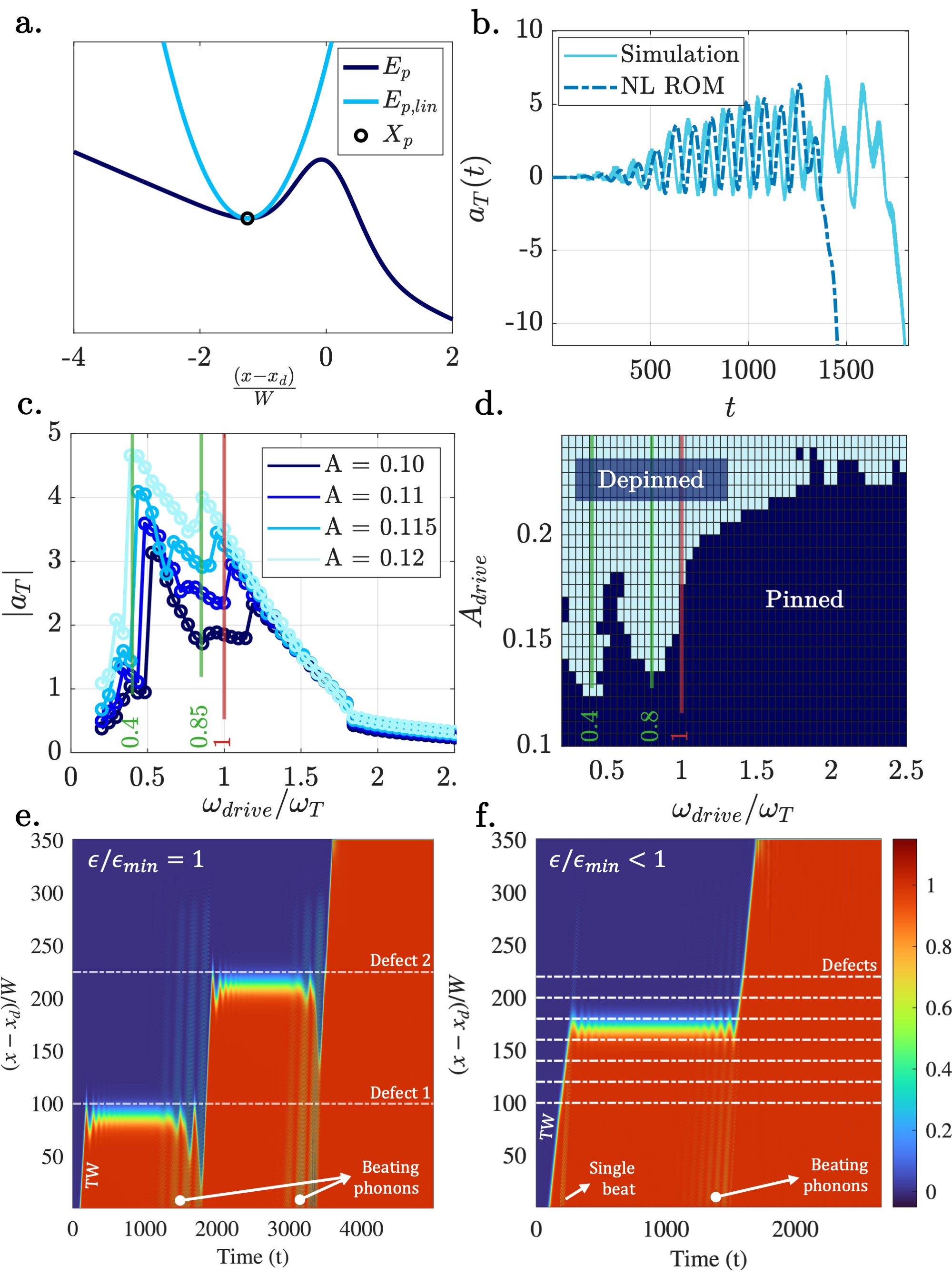}
\caption{\label{fig:fig3} (a) Harmonic potential of the linear ROM (light blue) compared to the potential energy of the kink quasi-particle (dark blue) (b) Comparison between simulation 
and non-linear ROM where high-amplitude excitations cause depinning. (c) Softening frequency response of the non-linear ROM for amplitudes just lower than depinning (0.125), showing two peaks at $0.4\omega_T$ and $0.85\omega_T$. (d) Simulations showing the frequency and amplitude dependence of depinning. (e) A system with two defects (at units 100 and 225), in which a TW is stopped and depinned twice. (f) System with array of defects ($\epsilon < \epsilon_{min}$); the TW passes through them until slowed down by a timed excitation (single beat), gets pinned, and subsequently depinned using beating phonons.}
\end{figure}

Depinning from the defect requires large motions of the kink in a strongly nonlinear potential ($E_p(X_k)$ from Eq.~\ref{eq:pot_en}), which cannot be accurately captured using a harmonic approximation of such potential. In fact, we see a strong softening of the dynamics (see SI Fig.~\ref{fig:SIsoftening}) when the excitation amplitude is increased. To account for this behavior, it's necessary to consider the position-dependent stiffness over large movements $X(t)$ of the kink (as shown by Fig.~\ref{fig:fig3}a).

Following the procedure described in SI Section~VIII, it is possible to demonstrate that $a_T(t) \approx -\sqrt{m_k} X(t)$, where $X(t) = X_k(t) - X_p$, $X_p$ is the position of the pinned kink center and $X_k(t)$ the time-dependent one. Using this relation and Eq. \ref{eq:pot_en}, the non-linear ROM can be written as:
\begin{equation}
    \ddot{a}_T + \beta \dot{a}_T + \frac{1}{\sqrt{m_k}} \frac{d E_p}{dX} = F_T(t).   
\end{equation}
For small $a_T$, this reduces to the previous linear ROM (Eq.~\ref{eq:a_j_ode_linear})
, with $\omega_T^2 = \frac{1}{m_k}\frac{d^2 E_p}{dX^2}(X_p)$.

This ROM not only captures the softening behavior of the oscillating kink, but it can also predict both depinning (see Fig.~\ref{fig:fig3}b) and the frequencies at which it can be achieved most efficiently. The frequency response in Fig.~\ref{fig:fig3}c shows two main peaks. The lower frequency one reflects the softening dynamics: as the oscillation amplitude grows, the effective natural frequency drops to about $0.4\omega_T$. The other peak occurs when the drive frequency ($\approx0.8\omega_T$) matches twice the softened natural frequency, indicating a superharmonic resonance. The depinning map in Fig.~\ref{fig:fig3}d (obtained using full lattice simulations) confirms that these are the frequencies at which the required forcing amplitudes are minimal. Similar optimal frequencies are obtained for different lattice asymmetries $\alpha$ using the semi-analytical procedure described in the SI Section~VIII and SI Fig.~\ref{fig:SIoptfreqpred}, showing that this ROM can be used as a predictive tool over a broad range of parameters.

Finally, we show how this mechanism operates in more complex settings. 
First, multiple defects are introduced, stopping the TW at different locations and then allowing it to propagate further (Fig.~\ref{fig:fig3}e). Phonons are not strongly reflected by defects with strength $\epsilon \approx \epsilon_{min}$ (Fig.~\ref{fig:SIRT}). This allows phonons to propagate through several defects without losing significant energy and repeatedly drive depinning, effectively demonstrating the feasibility of using phonons as a control strategy for multiple re-writing of memory states in the system. However, a major limitation is the need for relatively strong excitations when dissipation or system size increase (with beating amplitude $a_1+a_2\approx(1-\alpha)$ in extreme cases). Still, we find that consistent depinning is achievable for $\alpha>0.45$.
To mitigate this limitation, and expand the parameter space, we propose a variation to the current strategy, as illustrated in Fig~\ref{fig:fig3}f. Here, a defect with $\epsilon / \epsilon_{min}<1$ is introduced. Although it cannot independently stop the TW propagation, we exploit the static component of the force generated by the beating phonons (Fig.~\ref{fig:SIFt}), which effectively biases the forcing to pull the kink backwards, to slow the TW down and reduce its kinetic energy. This can be achieved with a short burst (of duration $\Delta t=1/\omega_T$) of beating phonons just before the kink interacts with the defect. Under these conditions, a sub-threshold defect can pin the kink. From here, depinning requires smaller forcing amplitudes and smaller oscillations of the kink.

We conclude with a few comments to provide general context for the strategy proposed in this work. 
First, despite being derived for a polynomial potential $V(u)$, the strategy described here inherently relies on the existence of a finite frequency translational-like mode within a band gap. Consequently, the results apply universally to the widely used sine-Gordon, $\phi4$ systems with external DC driving field, and similar models.

Furthermore, although we analyze a continuous system, the procedure can be extended to discrete lattices as well. While discreteness introduces a Peierls-Nabarro potential that modifies mode frequencies and effective dissipation, the nonlinear phonon-kink coupling fundamentally remains the same. For this reason, this strategy can be experimentally viable across different metamaterials, including systems based on bistable beams and springs (proposed in Fig. \ref{fig:fig1}a), Kane-Lubensky chains \cite{qian2025observation}, rotating square metamaterials, and more. The beating excitation can be implemented via a single shaker or a piezoelectric actuator at the boundary, reducing the complexity of current mechanical memory units and improving scalability and allowing possible miniaturization.

A critical advantage of our strategy for storing mechanical data as kinks is its robustness against external perturbations. In experimental settings, ambient vibrations typically span a broad spectrum. However, because the lattice prohibits wave propagation at the translational resonant frequency, the pinned kink is effectively shielded from random environmental noise, and depinning can be achieved only with specific forcing modulations. This makes the proposed strategy particularly viable and resilient for memory storage and mechanical computation in systems subjected to real world vibrations and external disturbances.

Finally, since phonons traveling through a kink experience a phase change \cite{hasenfratz1977interaction}, the location of the pinned kink can, in principle, be determined by measuring the phase change of the beating phonons from the actuation site to the opposite end of the system (see SI Section~VIII). This self-sensing could be relevant as a form of mechanical proprioception in mechanical computing systems~\cite{jiao2024proprioception}.

In summary, this paper examines the pinning and de-pinning of kink solitons in a 1D bistable lattice with an asymmetric potential and a localized defect. The threshold defect strength for pinning a propagating transition wave is derived using an approach that accounts for different damping regimes. 
We then established a general framework for the remote depinning of the kink using a pair of beating phonons that resonates with the kink's translational mode, bypassing the limitation that results from low-frequency band gaps. 
A non-linear ROM that accounts for the amplitude-dependent softening predicts the kink's oscillatory dynamics and predicts the optimal excitation frequency for efficient depinning. 
This mechanism is relevant to a broad range of bistable metamaterials that support transition waves, independent of specific architectures or scale. It provides a control strategy for kink solitons that can be applied to improve mechanical memory units with simpler, and more robust data storing.

\begin{acknowledgments}
The authors gratefully acknowledge support via AFOSR award numbers FA9550-23-1-0416 and FA9550-23-1-0299.
\end{acknowledgments}

\nocite{*}
\bibliography{bibliography}

@article{wang2023phase,
  title={Phase transitions in hierarchical, multi-stable metamaterials},
  author={Wang, Chongan and Frazier, Michael J},
  journal={Extreme Mechanics Letters},
  volume={64},
  pages={102068},
  year={2023},
  publisher={Elsevier}
}

@article{deng2022topological,
  title={Topological solitons make metamaterials crawl},
  author={Deng, Bolei and Zanaty, Mohamed and Forte, Antonio E and Bertoldi, Katia},
  journal={Physical Review Applied},
  volume={17},
  number={1},
  pages={014004},
  year={2022},
  publisher={APS}
}

@article{yasuda2021,
  title={Mechanical computing},
  author={Yasuda, H. and Buskohl, P. R. and Gillman, A. and Murphey, T. D. and Stepney, S. and Vaia, R. A. and Raney, J. R.},
  journal={Nature},
  volume={598},
  year={2021},
  pages={39-48}
}

@article{yasuda2023nucleation,
  title={Nucleation of transition waves via collisions of elastic vector solitons},
  author={Yasuda, Hiromi and Shu, Hang and Jiao, Weijian and Tournat, Vincent and Raney, Jordan R},
  journal={Applied Physics Letters},
  volume={123},
  number={5},
  pages={051701},
  year={2023},
  publisher={AIP Publishing}
}

@article{jiao2024nucleation,
  title={Phase transitions in 2D multistable mechanical metamaterials via collisions of soliton-like pulses},
  author={Jiao, W. and Shu, H. and Tournat, V. and Raney, J. R.},
  journal={Nature Communications},
  volume={15},
  pages={333},
  year={2024}
}

@article{song2019,
  title={Additively manufacturable micro-mechanical logic gates},
  author={Song, Y. and Panas, R. M. and Chizari, S. and Shaw, L. A. and Jackson, J. A. and Hopkins, J. B. and Pascall, A. J.},
  journal={Nature Communications},
  volume={10},
  year={2019},
  pages={882}
}

@article{jiang2019,
  title={Bifurcation-based embodied logic and autonomous actuation},
  author={Jiang, Y. and Korpas, L. M. and Raney, J. R.},
  journal={Nature Communications},
  volume={10},
  year={2019},
  pages={128}
}

@article{chen2025,
  title={General-purpose mechanical computing enabled by origami circuit reconfiguration with robotic addressing and activation},
  author={Chen, Y. and Tan, T. and Yan, Z.},
  journal={Nature Communications},
  volume={16},
  year={2025},
  pages={10483}
}

@article{jiao2024proprioception,
  title={Toward mechanical proprioception in autonomously reconfigurable kirigami-inspired mechanical systems},
  author={Jiao, W. and Shu, H. and He, Q. and Raney, J. R.},
  journal={Philosophical Transactions of the Royal Society A},
  volume={382},
  number={2283},
  year={2024},
  pages={20240116}
}

@article{deng2021nonlinear,
  title={Nonlinear waves in flexible mechanical metamaterials},
  author={Deng, Bolei and Raney, Jordan R and Bertoldi, Katia and Tournat, Vincent},
  journal={Journal of Applied Physics},
  volume={130},
  number={4},
  year={2021},
  publisher={AIP Publishing}
}

@article{malomed1992perturbative,
  title={Perturbative analysis of the interaction of a phi 4 kink with inhomogeneities},
  author={Malomed, BA},
  journal={Journal of Physics A: Mathematical and General},
  volume={25},
  number={4},
  pages={755},
  year={1992},
  publisher={IOP Publishing}
}

@article{qian2025observation,
  title={Observation of mechanical kink control and generation via phonons},
  author={Qian, Kai and Cheng, Nan and Serafin, Francesco and Sun, Kai and Theocharis, Georgios and Mao, Xiaoming and Boechler, Nicholas},
  journal={arXiv preprint arXiv:2502.16117},
  year={2025}
}

@article{cuevas2014sine,
  title={The sine-Gordon model and its applications},
  author={Cuevas-Maraver, Jes{\'u}s and Kevrekidis, Panayotis G and Williams, Floyd},
  journal={Nonlinear systems and complexity},
  volume={10},
  year={2014},
  publisher={Springer}
}

@article{chen2017effects,
  title={Effects of phonons on mobility of dislocations and dislocation arrays},
  author={Chen, Xiang and Xiong, Liming and McDowell, David L and Chen, Youping},
  journal={Scripta Materialia},
  volume={137},
  pages={22--26},
  year={2017},
  publisher={Elsevier}
}

@article{shilo2003stroboscopic,
  title={Stroboscopic X-ray imaging of vibrating dislocations excited by 0.58 GHz phonons},
  author={Shilo, D and Zolotoyabko, E},
  journal={Physical review letters},
  volume={91},
  number={11},
  pages={115506},
  year={2003},
  publisher={APS}
}

@article{siu2011understanding,
  title={Understanding acoustoplasticity through dislocation dynamics simulations},
  author={Siu, KW and Ngan, AHW},
  journal={Philosophical Magazine},
  volume={91},
  number={34},
  pages={4367--4387},
  year={2011},
  publisher={Taylor \& Francis}
}

@article{swinburne2013theory,
  title={Theory and simulation of the diffusion of kinks on dislocations in bcc metals},
  author={Swinburne, TD and Dudarev, SL and Fitzgerald, SP and Gilbert, MR and Sutton, AP},
  journal={Physical Review B—Condensed Matter and Materials Physics},
  volume={87},
  number={6},
  pages={064108},
  year={2013},
  publisher={APS}
}

@article{hasenfratz1977interaction,
  title={The interaction of a solitary wave solution with phonons in a one-dimensional model for displacive structural phase transitions},
  author={Hasenfratz, Wolfgang and Klein, R},
  journal={Physica A: Statistical Mechanics and its Applications},
  volume={89},
  number={1},
  pages={191--204},
  year={1977},
  publisher={Elsevier}
}

@article{pechac2023mechanical,
  title={Mechanical multi-level memory from multi-stable metamaterial},
  author={Pechac, Jack E and Frazier, Michael J},
  journal={Applied Physics Letters},
  volume={122},
  number={21},
  year={2023},
  publisher={AIP Publishing}
}

@article{deng2020characterization,
  title={Characterization, stability, and application of domain walls in flexible mechanical metamaterials},
  author={Deng, Bolei and Yu, Siqin and Forte, Antonio E and Tournat, Vincent and Bertoldi, Katia},
  journal={Proceedings of the National Academy of Sciences},
  volume={117},
  number={49},
  pages={31002--31009},
  year={2020},
  publisher={National Academy of Sciences}
}

@article{jin2020guided,
  title={Guided transition waves in multistable mechanical metamaterials},
  author={Jin, Lishuai and Khajehtourian, Romik and Mueller, Jochen and Rafsanjani, Ahmad and Tournat, Vincent and Bertoldi, Katia and Kochmann, Dennis M},
  journal={Proceedings of the National Academy of Sciences},
  volume={117},
  number={5},
  pages={2319--2325},
  year={2020},
  publisher={National Academy of Sciences}
}

@article{wang2023phase2,
  title={Phase patterning in multi-stable metamaterials: Transition wave stabilization and mode conversion},
  author={Wang, Chongan and Frazier, Michael J},
  journal={Applied Physics Letters},
  volume={123},
  number={1},
  year={2023},
  publisher={AIP Publishing}
}

@article{tahidul2024reprogrammable,
  title={Reprogrammable mechanics via individually switchable bistable unit cells in a prestrained chiral metamaterial},
  author={Tahidul Haque, ABM and Ferracin, Samuele and Raney, Jordan R},
  journal={Advanced Materials Technologies},
  volume={9},
  number={17},
  pages={2400474},
  year={2024},
  publisher={Wiley Online Library}
}

@article{dudek2025shape,
  title={Shape-morphing metamaterials},
  author={Dudek, Krzysztof K and Kadic, Muamer and Coulais, Corentin and Bertoldi, Katia},
  journal={Nature Reviews Materials},
  pages={1--16},
  year={2025},
  publisher={Nature Publishing Group UK London}
}

@article{liang2025ideal,
  title={Ideal energy-absorbing metamaterials based on self-locking bistable structures},
  author={Liang, Kuan and Zhang, Xiaopeng and Zhao, Qi and Suo, Liujia and Wei, Zishen and Wang, Yaguang and Luo, Yangjun and Takezawa, Akihiro and Wang, Dazhi},
  journal={Materials Horizons},
  volume={12},
  number={12},
  pages={4165--4176},
  year={2025},
  publisher={Royal Society of Chemistry}
}

@article{raney2016stable,
  title={Stable propagation of mechanical signals in soft media using stored elastic energy},
  author={Raney, Jordan R and Nadkarni, Neel and Daraio, Chiara and Kochmann, Dennis M and Lewis, Jennifer A and Bertoldi, Katia},
  journal={Proceedings of the National Academy of Sciences},
  volume={113},
  number={35},
  pages={9722--9727},
  year={2016},
  publisher={National Academy of Sciences}
}

@article{nadkarni2016unidirectional,
  title={Unidirectional transition waves in bistable lattices},
  author={Nadkarni, Neel and Arrieta, Andres F and Chong, Christopher and Kochmann, Dennis M and Daraio, Chiara},
  journal={Physical review letters},
  volume={116},
  number={24},
  pages={244501},
  year={2016},
  publisher={APS}
}

@article{peyrard1984kink,
  title={Kink dynamics in the highly discrete sine-Gordon system},
  author={Peyrard, Michel and Kruskal, Martin D},
  journal={Physica D: Nonlinear Phenomena},
  volume={14},
  number={1},
  pages={88--102},
  year={1984},
  publisher={Elsevier}
}

@article{currie1979dynamics,
  title={Dynamics of domain walls in ferrodistortive materials II. Applications to Pb 5 Ge 3 O 11-and SbSI-type ferroelectrics},
  author={Currie, JF and Blumen, A and Collins, MA and Ross, John},
  journal={Physical Review B},
  volume={19},
  number={7},
  pages={3645},
  year={1979},
  publisher={APS}
}

@article{puglisi2000mechanics,
  title={Mechanics of a discrete chain with bi-stable elements},
  author={Puglisi, G and Truskinovsky, Lev},
  journal={Journal of the Mechanics and Physics of Solids},
  volume={48},
  number={1},
  pages={1--27},
  year={2000},
  publisher={Elsevier}
}

@book{abeyaratne2006evolution,
  title={Evolution of phase transitions: a continuum theory},
  author={Abeyaratne, Rohan and Knowles, James K},
  year={2006},
  publisher={Cambridge University Press}
}

@article{malomed1993interactions,
  title={Interactions of kinks with defect modes},
  author={Malomed, Boris A and Campbell, David K and Knowles, Noah and Flesch, Randy J},
  journal={Physics Letters A},
  volume={178},
  number={3-4},
  pages={271--278},
  year={1993},
  publisher={Elsevier}
}

@article{fei1992resonant,
  title={Resonant kink-impurity interactions in the $\varphi$ 4 model},
  author={Fei, Zhang and Kivshar, Yuri S and V{\'a}zquez, Luis},
  journal={Physical Review A},
  volume={46},
  number={8},
  pages={5214},
  year={1992},
  publisher={APS}
}

@article{he2024programmable,
  title={Programmable responsive metamaterials for mechanical computing and robotics},
  author={He, Qiguang and Ferracin, Samuele and Raney, Jordan R},
  journal={Nature Computational Science},
  volume={4},
  number={8},
  pages={567--573},
  year={2024},
  publisher={Nature Publishing Group US New York}
}

@article{chen2021reprogrammable,
  title={A reprogrammable mechanical metamaterial with stable memory},
  author={Chen, Tian and Pauly, Mark and Reis, Pedro M},
  journal={Nature},
  volume={589},
  number={7842},
  pages={386--390},
  year={2021},
  publisher={Nature Publishing Group UK London}
}

@article{mei2021mechanical,
  title={A mechanical metamaterial with reprogrammable logical functions},
  author={Mei, Tie and Meng, Zhiqiang and Zhao, Kejie and Chen, Chang Qing},
  journal={Nature communications},
  volume={12},
  number={1},
  pages={7234},
  year={2021},
  publisher={Nature Publishing Group UK London}
}

@article{byun2024integrated,
  title={Integrated mechanical computing for autonomous soft machines},
  author={Byun, Junghwan and Pal, Aniket and Ko, Jongkuk and Sitti, Metin},
  journal={Nature Communications},
  volume={15},
  number={1},
  pages={2933},
  year={2024},
  publisher={Nature Publishing Group UK London}
}

@article{watkins2025arbitrary,
  title={Arbitrary mechanical memory encoding via nonlinear waves in bistable metamaterials},
  author={Watkins, Audrey A and Bordiga, Giovanni and Mu, Mingxing and Tournat, Vincent and Bertoldi, Katia},
  journal={arXiv preprint arXiv:2508.20321},
  year={2025}
}

@article{braun1998nonlinear,
  title={Nonlinear dynamics of the Frenkel--Kontorova model},
  author={Braun, Oleg M and Kivshar, Yuri S},
  journal={Physics Reports},
  volume={306},
  number={1-2},
  pages={1--108},
  year={1998},
  publisher={Elsevier}
}

@article{paliovaios2024transition,
  title={Transition waves in bistable systems generated by collision of moving breathers},
  author={Paliovaios, A and Theocharis, Georgios and Achilleos, V and Tournat, V},
  journal={Extreme Mechanics Letters},
  volume={71},
  pages={102199},
  year={2024},
  publisher={Elsevier}
}

@article{yasuda2017origami,
  title={Origami-based tunable truss structures for non-volatile mechanical memory operation},
  author={Yasuda, Hiromi and Tachi, Tomohiro and Lee, Mia and Yang, Jinkyu},
  journal={Nature communications},
  volume={8},
  number={1},
  pages={962},
  year={2017},
  publisher={Nature Publishing Group UK London}
}

@article{jules2022delicate,
  title={Delicate memory structure of origami switches},
  author={Jules, Th{\'e}o and Reid, Austin and Daniels, Karen E and Mungan, Muhittin and Lechenault, Fr{\'e}d{\'e}ric},
  journal={Physical Review Research},
  volume={4},
  number={1},
  pages={013128},
  year={2022},
  publisher={APS}
}

@article{stenseng2025bi,
  title={Bi-Stable Metamaterials with Intrinsic Memory for Selective Wave Filtering Based on Frequency and Amplitude},
  author={Stenseng, Nathan N and Samak, Mahmoud M and Bilal, Osama R},
  journal={Advanced Science},
  volume={12},
  number={1},
  pages={2405146},
  year={2025},
  publisher={Wiley Online Library}
}

@article{bodaghi20194d,
  title={4D printed tunable mechanical metamaterials with shape memory operations},
  author={Bodaghi, M and Liao, WH},
  journal={Smart Materials and Structures},
  volume={28},
  number={4},
  pages={045019},
  year={2019},
  publisher={IOP Publishing}
}

@article{liu2023cellular,
  title={Cellular automata inspired multistable origami metamaterials for mechanical learning},
  author={Liu, Zuolin and Fang, Hongbin and Xu, Jian and Wang, Kon-Well},
  journal={Advanced Science},
  volume={10},
  number={34},
  pages={2305146},
  year={2023},
  publisher={Wiley Online Library}
}


\clearpage
\onecolumngrid

\clearpage
\onecolumngrid

\counterwithout{equation}{section}
\counterwithout{figure}{section}
\counterwithout{table}{section}

\setcounter{equation}{0}
\setcounter{figure}{0}
\setcounter{table}{0}
\setcounter{page}{1}

\renewcommand{\theequation}{S\arabic{equation}}
\renewcommand{\thefigure}{S\arabic{figure}}
\renewcommand{\thetable}{S\arabic{table}}

\begin{center}
  \textbf{\large Supplementary Information: Phonon controlled mechanical memory\\via pinning and depinning of transition waves}\\[.2cm]
  Samuele Ferracin, Dengge Jin, Vincent Tournat, Jordan Raney\\[.5cm]
\end{center}



\section{Governing Equation}

\begin{figure}[h]
\centering
\includegraphics[width=\textwidth]{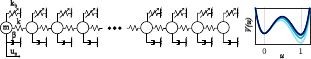}
\caption{\label{fig:SIfigEe} Schematics of the discrete system from Eq. \ref{eq:SIinitialeq}.}
\end{figure}

The starting governing equations of our discrete system are $\forall n$:

\begin{equation}\label{eq:SIinitialeq}
m \frac{d^2 u_n}{dt^2} + \beta \frac{d u_n}{dt} = k (u_{n-1} - u_n) + k (u_{n+1} - u_n) - k_b u_n (u_n - \alpha L)(u_n - L),
\end{equation}
where $u_n$ is the displacement of unit $n$, $m$ its mass, $\beta$ the viscous damping parameter. The masses are connected to their first neighbor with linear springs of stiffness $k$ [$N/m$] and are grounded with bistable springs of stiffness $k_b$ [$N/m^3$] with cubic nonlinearity. These nonlinear springs can be asymmetric, where the parameter $0<\alpha<1$ governs the asymmetry.

\subsection{Continuum approximation}
We set \( x = n a \), so \( u_n(t) = u(n a, t) \), where $a$ is the lattice spacing:

\[
u_{n+1}(t) = u(x + a, t) = u(x, t) + a \frac{\partial u}{\partial x} + \frac{a^2}{2} \frac{\partial^2 u}{\partial x^2} + \frac{a^3}{6} \frac{\partial^3 u}{\partial x^3} + O(a^4),
\]
\[
u_{n-1}(t) = u(x - a, t) = u(x, t) - a \frac{\partial u}{\partial x} + \frac{a^2}{2} \frac{\partial^2 u}{\partial x^2} - \frac{a^3}{6} \frac{\partial^3 u}{\partial x^3} + O(a^4),
\]
\[
u_{n+1} + u_{n-1} - 2 u = a^2 \frac{\partial^2 u}{\partial x^2} + O(a^4).
\]

Neglecting the higher-order terms:
\[
\frac{\partial^2 u}{\partial t^2} = - \frac{\beta}{m} \frac{\partial u}{\partial t} + \frac{k a^2}{m} \frac{\partial^2 u}{\partial x^2} - \frac{k_b}{m} u (u - \alpha L)(u - L),
\]

\begin{equation}
\frac{\partial^2 u}{\partial t^2} + \frac{\beta}{m} \frac{\partial u}{\partial t} - c^2 \frac{\partial^2 u}{\partial x^2} + \frac{k_b}{m} u (u - \alpha L)(u - L) = 0,
\end{equation}
with
\[
c^2 = \frac{k a^2}{m}.
\]

\subsection{Dimensionless Variables}
\begin{itemize}
    \item \(\bar{u} = \frac{u}{a}\), where \(u\) is the displacement, and where \(a\) is the lattice spacing between masses. We consider, from here, on $L=a$ for convenience;
    \item \(\bar{x} = \frac{x}{a}\);
    \item \(\bar{t} = \frac{t}{T}\), where \(T\) is a characteristic time scale.
\end{itemize}

Then:
\[
\frac{\partial u}{\partial x} = \frac{\partial \bar{u}}{\partial \bar{x}}, \quad \frac{\partial^2 u}{\partial x^2} = \frac{1}{a} \frac{\partial^2 \bar{u}}{\partial \bar{x}^2},
\]

\[
\frac{\partial u}{\partial t} = \frac{a}{T} \frac{\partial \bar{u}}{\partial \bar{t}}, \quad \frac{\partial^2 u}{\partial t^2} = \frac{a}{T^2} \frac{\partial^2 \bar{u}}{\partial \bar{t}^2}.
\]

Substituting in the equation of motion:

\[
\frac{a}{T^2} \frac{\partial^2 \bar{u}}{\partial \bar{t}^2} + \frac{a}{T} \frac{\beta}{m} \frac{\partial \bar{u}}{\partial \bar{t}} - \frac{c^2 }{a} \frac{\partial^2 \bar{u}}{\partial \bar{x}^2} + \frac{k_b}{m} (\bar{u} a) (\bar{u} a - \alpha a)(\bar{u} a - a) = 0.
\]

Dividing by \( \frac{a}{T^2} \) we get the non dimensional quantities:

\[
\frac{\partial^2 \bar{u}}{\partial \bar{t}^2} + T \frac{\beta}{m} \frac{\partial \bar{u}}{\partial \bar{t}}  - c^2 \frac{T^2}{a^2} \frac{\partial^2 \bar{u}}{\partial \bar{x}^2} + \frac{k_b}{m} T^2 a^2 \bar{u} \left( \bar{u} - \alpha \right) \left( \bar{u} - 1 \right) = 0,
\]

\[
\frac{k_b}{m} T^2 a^2 = 1, \quad T^2 = \frac{m}{k_b a^2},
\]

\[
\frac{c^2 T^2}{a^2} = \frac{k a^2}{m} \cdot \frac{m}{k_b a^2} \cdot \frac{1}{a^2} = \frac{k}{k_b a^2} = \bar{c}^2,
\]

\[
T \frac{\beta}{m}= \sqrt{\frac{m}{k_b a^2}}\frac{\beta}{m} =\sqrt{\frac{1}{k_b a^2}}\beta=\bar{\beta}.
\]

Finally:
\begin{equation}\label{eq:SInondimeq1}
\ddot{\bar{u}}+ \bar{\beta}\dot{\bar{u}} - \bar{c}^2 \bar{u}_{,\bar{x}\bar{x}} + \bar{u} \left( \bar{u} - \alpha \right) \left( \bar{u} - 1 \right) = 0,
\end{equation}
\[
\ddot{\bar{u}} + \bar{\beta}\dot{\bar{u}} - \bar{c}^2 \bar{u}_{,\bar{x}\bar{x}} + \bar{u}^3 - (\alpha + 1) \bar{u}^2 + \alpha \bar{u} = 0,
\]

\begin{equation}\label{eq:SInondimeq2}
\ddot{\bar{u}} + \bar{\beta}\dot{\bar{u}} - \bar{c}^2 \bar{u}_{,\bar{x}\bar{x}} + \frac{dV}{d\bar{u}} = 0.
\end{equation}

This is a Klein-Gordon equation with cubic nonlinearity \( \frac{dV}{d\bar{u}} = \bar{u}^3 - (\alpha + 1) \bar{u}^2 + \alpha \bar{u} \) , where $V(\bar{u})$ is the nonlinear bistable potential. We can integrate the nonlinear force to find the potential:
\begin{equation}\label{eq:SIpotential}
V(\bar{u}) = \int (\bar{u}^3 - (\alpha + 1) u^2 + \alpha \bar{u}) \, d\bar{u} = \frac{1}{4} \bar{u}^4 - \frac{\alpha + 1}{3} \bar{u}^3 + \frac{\alpha}{2} \bar{u}^2
\end{equation}
(setting the integration constant to zero). The roots are \( \bar{u} = 0 \)  ( $V(\bar{u}=0)=0$ ), \( \bar{u} = 1 \) ( $V(\bar{u}=1)=\frac{2\alpha-1}{12}$ ) and \( \bar{u} = \alpha \). For \( \alpha \in (0, 1) \), stable equilibria are at \( \bar{u} = 0 \) and \( \bar{u} = 1 \), and the unstable equilibrium is at \( \bar{u} = \alpha \).
\\
From here onward, we abandon the $\bar{}$ notation to address the non dimensional units.

\section{Linear waves}
From this governing equation, in an undamped scenario, we consider small amplitudes around $u=0$, and $u=1$. The nonlinear force is linearized as $F(u)\approx-\alpha u$ in the first case, and $F(u)\approx-(1-\alpha)u$ in the second.

\paragraph{Continuum limit}
We assume a wave solution of the form
\begin{equation}
\bar{u}=Ae^{i(qx-\omega t)},
\end{equation}
with wavenumber $q_0$ and $q_1$, depending which equilibrium state is considered.
This gives:
\begin{equation}\label{eq:linearwaves}
\begin{split}
    \omega_0^2=c^2q_0^2+\alpha,
    \\
    \omega_1^2=c^2q_1^2+1-\alpha,
\end{split}
\end{equation}
which are classic Klein-Gordon dispersions \cite{braun1998nonlinear} (see Fig. \ref{fig:fig2}a) with a low frequency bandgap, which cutoff frequency $\omega_{cutoff}$ depends on the asymmetry parameter $\alpha$ and the equilibrium state.
The phase velocity is, as expected, always $\geq c$:
\begin{equation}
\begin{split}
    v_{p0}(q_0) = \frac{\omega_0}{q_0}= \sqrt{c^2+\frac{\alpha}{q_0^2}},
    \\
    v_{p1}(q_1) = \frac{\omega_1}{q_1}= \sqrt{c^2+\frac{1-\alpha}{q_1^2}}.
\end{split} 
\end{equation}
The group velocity is:
\begin{equation}
\begin{split}
    v_{g0}(q_0) = \frac{d\omega_0}{dq_0}= \frac{1}{2}\frac{2c^2q_0}{\sqrt{c^2q_0^2+\alpha}}=\frac{c^2q_0}{\omega_0},
    \\
    v_{g1}(q_1) = \frac{d\omega_1}{dq_1}= \frac{1}{2}\frac{2c^2q_1}{\sqrt{c^2q_1^2+1-\alpha}}=\frac{c^2q_1}{\omega_1}.
\end{split}
\end{equation}

\paragraph{Discrete}
We assume a wave solution of the form:
\begin{equation}
u_j=Ae^{i(qja_s-\omega t)},
\end{equation}
where $a_s$ is the non-dimensional spacing length chosen to numerically simulate the system. The spacing is chosen to be small enough ($a_s=W/10$) to maintain valid the continuum approximation. Substituting this solution in the discrete equation
\begin{equation}
    u_{j,tt} - \frac{c^2}{a_s^2} (u_{j+1}-2u_{j}+u_{j-1}) - F(u) = 0,
\end{equation}
gives:
\begin{equation}\label{eq:linearwavesdiscrete}
\begin{split}
    \omega_0^2=\frac{2c^2}{a_s^2}(1-\cos(q_0a_s))+\alpha,
    \\
    \omega_1^2=\frac{2c^2}{a_s^2}(1-\cos(q_1a_s))+1-\alpha.
\end{split}
\end{equation}
The group velocity is:
\begin{equation}\label{eq:SIdiscretedispersion}
\begin{split}
    v_{g0}(q_0) = \frac{d\omega_0}{dq_0}=\frac{2c^2}{\omega_0a_s}\sin(q_0a_s) ,
    \\
    v_{g1}(q_1) = \frac{d\omega_1}{dq_1}= \frac{2c^2}{\omega_1a_s}\sin(q_1a_s) .
\end{split}
\end{equation}

\newpage
\section{Transition wave analytical solution}
Here we follow the procedure in Section S3 of \cite{wang2023phase} to analytically find the form of the propagating transition wave.

From Eq. \ref{eq:SInondimeq2}:
\[
\ddot{u}  + \beta \dot{u}- c^2 u_{,xx} + \frac{dV}{du} = 0,
\]
we assume a traveling wave: \( u(x, t) = u(z) \), where
\[
z = \frac{x - v t}{\sqrt{1 - v^2}}.
\]
The derivatives are: \( \dot{u} = -\frac{v}{\sqrt{1 - v^2}} \frac{du}{dz} \), \( \ddot{u} = \left( \frac{v}{\sqrt{1 - v^2}} \right)^2 \frac{d^2 u}{dz^2} \), \( u_{,xx} = \left( \frac{1}{\sqrt{1 - v^2}} \right)^2 \frac{d^2 u}{dz^2} \).
\\Substituting in the equation of motion:
\[
\frac{(c^2-v^2)}{1 - v^2} \frac{d^2 u}{dz^2} + \beta \frac{v}{\sqrt{1 - v^2}} \frac{du}{dz} - \frac{dV}{du} = 0.
\]
Now we can reduce to a first order equation by a change of variable: \( p = \frac{du}{dz} \), so \( \frac{d^2 u}{dz^2} = p \frac{dp}{du} \):
\[
\frac{(c^2-v^2)}{1 - v^2} p \frac{dp}{du} + \beta \frac{v}{\sqrt{1 - v^2}} p = u^3 - (\alpha + 1) u^2 + \alpha u.
\]
\\
$p(u)$ has to be a second order polynomial, since the nonlinear force is a cubic one. Since the fixed point are $u=0$, and $u=1$, we can hypothesize $p(u)$ in the form:
\[ p = A u (u - 1), \quad \frac{dp}{du} = A (2u - 1). \]
Substituting:
\[
\frac{(c^2-v^2)}{1 - v^2} A^2 u (u - 1) (2u - 1) + \beta \frac{v}{\sqrt{1 - v^2}} A u (u - 1) = u^3 - (\alpha + 1) u^2 + \alpha u = u(u-\alpha)(u-1).
\]
We can factor out \( u (u - 1) \):
\[
\frac{(c^2-v^2)}{1 - v^2} A^2 (2u - 1) + \beta \frac{v}{\sqrt{1 - v^2}} A = u - \alpha.
\]
Equating the coefficients of the same orders: 
\[ 2 \frac{(c^2-v^2)}{1 - v^2} A^2 = 1, \quad A = \sqrt{\frac{1-v^2}{2(c^2-v^2)}}, \]
\[
v = \frac{c |1 - 2\alpha|}{\sqrt{2\beta^2 + (1 - 2\alpha)^2}}.
\]

The traveling speed $v$ of the TW depends on the asymmetry parameter $\alpha$ and on the damping $\beta$. Fig. \ref{fig:SIv0} shows the trend of this velocity for different $\beta$ and $\alpha$ values; when damping is very small, the propagation speed of the kink tends to $c$. We can also note that, for finite damping, the closer $\alpha$ is to 0, the faster the TW will be. If $alpha>0.5$, the asymmetry is reversed and the TW can propagate in the negative direction; if $\alpha=0.5$, and in the presence of damping, the TW will slow down and stop.
Substituting $A$ in $p$: \( \frac{du}{dz} = \sqrt{\frac{1-v^2}{2(c^2-v^2)}} u (u - 1) \) and separating variables:
\[
\frac{du}{u (u - 1)} = \sqrt{\frac{1-v^2}{2(c^2-v^2)}} dz.
\]
Using the partial fractions: \( \frac{1}{u (u - 1)} = \frac{1}{u - 1} - \frac{1}{u} \), we can integrate as:
\[
\ln \left| \frac{u - 1}{u} \right| = \sqrt{\frac{1-v^2}{2(c^2-v^2)}} (z - z_0),
\]
\[
\frac{u - 1}{u} = e^{\frac{(z - z_0)\sqrt{1-v^2}}{\sqrt{2(c^2-v^2)}}},
\]
\[
u(z) = \frac{1}{1 + e^{\frac{(z - z_0)\sqrt{1-v^2}}{\sqrt{2(c^2-v^2)}}}}.
\]

We transform \( u(z) = \frac{1}{1 + e^{\frac{(z - z_0)\sqrt{1-v^2}}{\sqrt{2(c^2-v^2)}}}} \) using the identity \( e^{x} = \frac{1 + \tanh(x/2)}{1 - \tanh(x/2)} \), where \( x = \frac{(z - z_0)\sqrt{1-v^2}}{\sqrt{2(c^2-v^2)}} \), so:
\[
e^{\frac{(z - z_0)\sqrt{1-v^2}}{\sqrt{2(c^2-v^2)}}} = \frac{1 + \tanh\left(\frac{(z - z_0)\sqrt{1-v^2}}{\sqrt{2(c^2-v^2)}}\right)}{1 - \tanh\left(\frac{(z - z_0)\sqrt{1-v^2}}{\sqrt{2(c^2-v^2)}}\right)},
\]
Substituting:
\[
u(z) = \frac{1 - \tanh\left(\frac{(z - z_0)\sqrt{1-v^2}}{\sqrt{2(c^2-v^2)}}\right)}{2}.
\]
We can obtain the analytical form of the transition wave:
\begin{equation}\label{eq:SItwanalytical}
u(z) = \frac{1}{2} \left[ 1 - \tanh\left( \frac{(z - z_0)\sqrt{1-v^2}}{\sqrt{2(c^2-v^2)}} \right) \right] = u(x,t) = \frac{1}{2} \left[ 1 - \tanh\left( \frac{(x - x_0) - v (t-t_0)}{W} \right) \right],
\end{equation}

\begin{equation}\label{eq:SIv0}
v = \sqrt{\frac{c^2 (1 - 2\alpha)^2}{2\beta^2 + (1 - 2\alpha)^2}},
\end{equation}
\begin{equation}\label{eq:SIW}
W = 2\sqrt{2(c^2-v^2)}.
\end{equation}
This is the classic tanh profile of a transition wave, propagating with speed $v$ and with a characteristic non dimensional width $W$. The latter dependence on $\beta$ and $\alpha$ can be seen in Fig. \ref{fig:SIWidth}: the profile becomes steeper for small damping (high velocities), tending to a step function. For high damping, instead, the width converges to the finite value $W=2c\sqrt{2}$.

\subsubsection{Analytical expressions of the kinetic and elastic energies of the kink}
The kinetic energy is:
\begin{equation}
E_k=\frac{1}{2}\int_{-\infty}^{\infty} \dot{u}^2 \,dx,
\end{equation}
and the elastic energy is:
\begin{equation}
E_e=\frac{c^2}{2}\int_{-\infty}^{\infty} u_{,x}^2 \,dx.
\end{equation}
Changing variable from $x$ to $\xi=\frac{x-vt}{W}$:
\[
\frac{d\xi}{dx}=\frac{1}{W},
\]
\[
u_{,t}=\frac{du}{d\xi}\frac{d\xi}{dt}=-\frac{1}{2}\mbox{sech}^2(\xi)(-\frac{v}{W}),
\]
\[
u_{,x}=\frac{du}{d\xi}\frac{d\xi}{dx}=-\frac{1}{2}\mbox{sech}^2(\xi)(\frac{1}{W}).
\]
Substituting these into the energy relations, the kinetic energy becomes:
\begin{equation}\label{eq:SIEk}
E_k=\frac{1}{2}\int_{-\infty}^{\infty} \frac{1}{4}\mbox{sech}^4(\xi)\frac{v^2}{W^2} \,Wd\xi=\frac{v^2}{8W}\frac{4}{3}=\frac{v^2}{6W},
\end{equation}
and the elastic energy:
\begin{equation}\label{eq:SIEe}
E_e=\frac{c^2}{2}\int_{-\infty}^{\infty} \frac{1}{4}\mbox{sech}^4(\xi)\frac{1}{W^2} \,Wd\xi=\frac{c^2}{8W}\frac{4}{3}=\frac{c^2}{6W}.
\end{equation}

\begin{figure}
\centering
\includegraphics[width=0.6\textwidth]{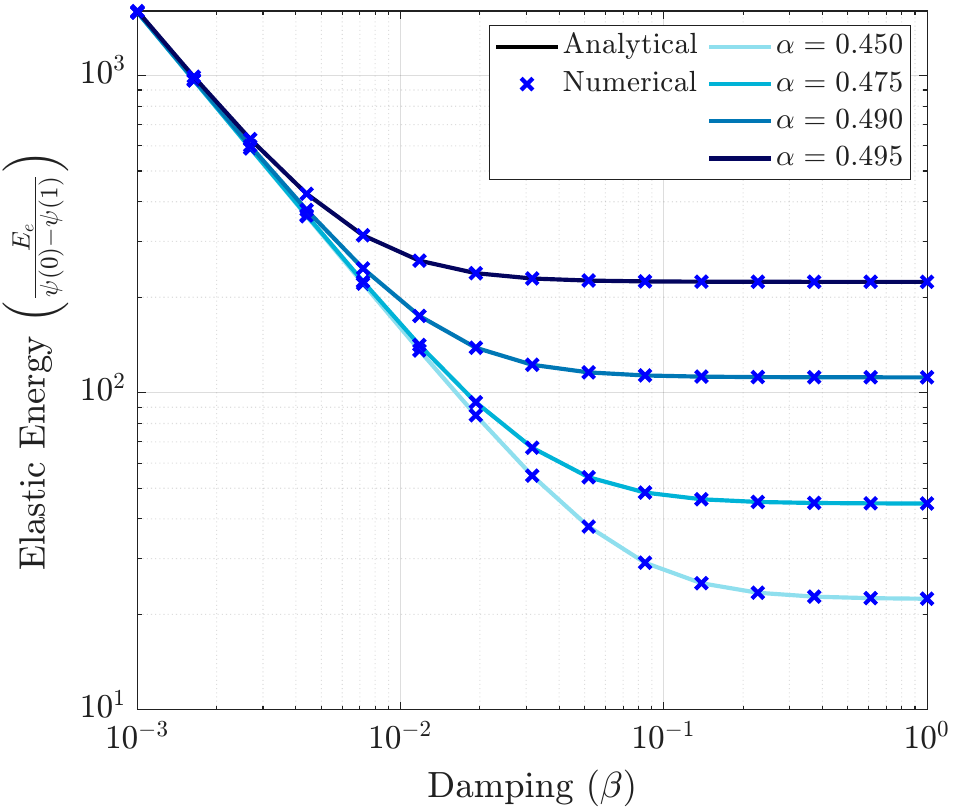}
\caption{\label{fig:SIfigEe} Elastic energy, analytical form from Eq. \ref{eq:SIEe} compared to the full simulation (using Matlab's \textit{ode45}). Here it's possible to see how, by refining the simulations with higher densities of points (reducing the spacing $a_s$), as explained before in SI Section 2, the results always match well the simulations results, without the difference that a strongly discrete system would introduce.}
\end{figure}
\begin{figure}
\centering
\includegraphics[width=0.6\textwidth]{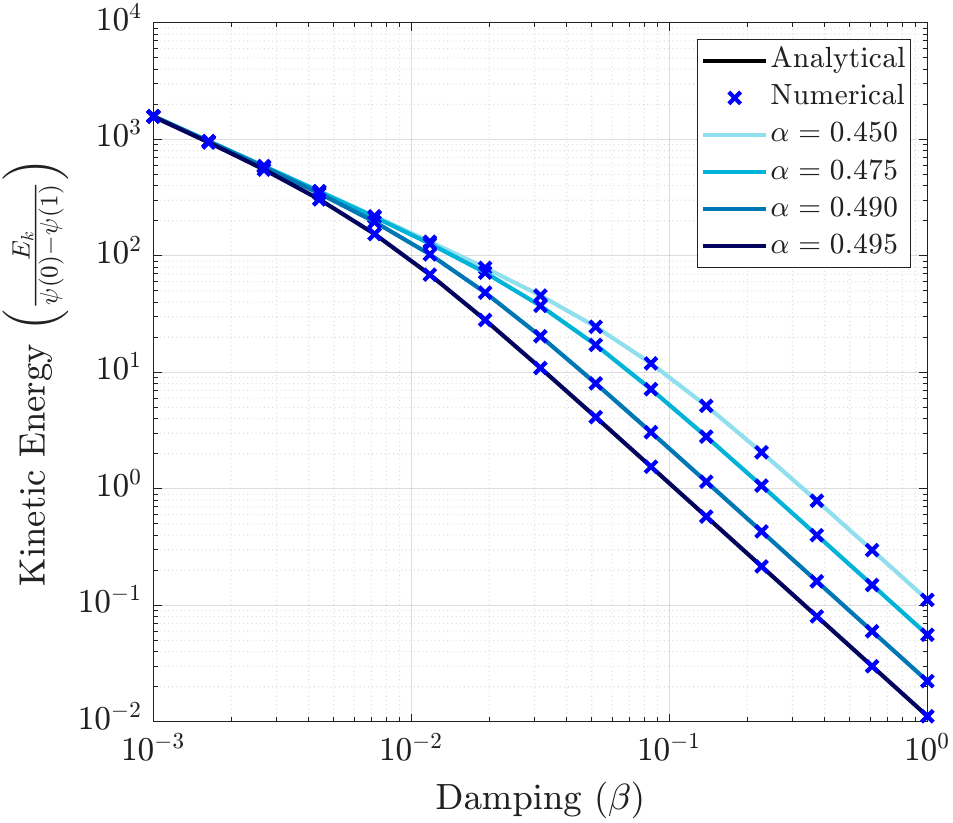}
\caption{\label{fig:SIfigEk} Kinetic energy, analytical form from Eq. \ref{eq:SIEk} compared to the full simulation (using Matlab's \textit{ode45}).}
\end{figure}

\begin{figure}
\centering
\includegraphics[width=0.6\textwidth]{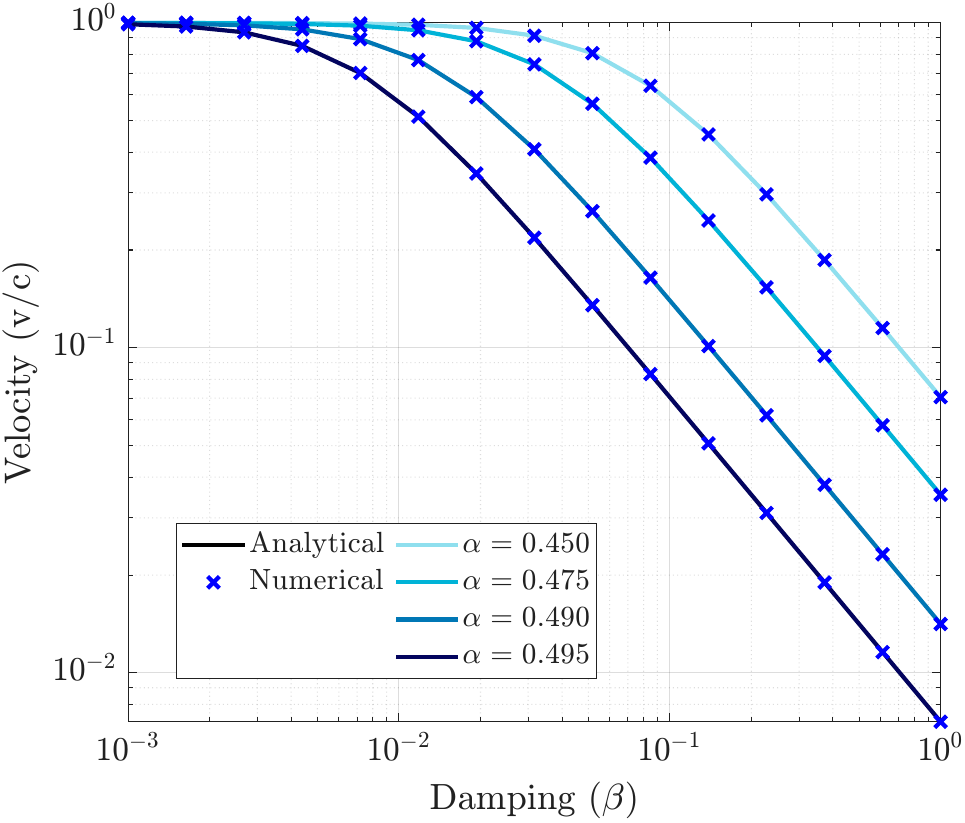}
\caption{\label{fig:SIv0} Velocity $v$ of the propagating kink, analytical form from Eq. \ref{eq:SIv0} compared to the full simulation (using Matlab's \textit{ode45}).}
\end{figure}
\begin{figure}
\centering
\includegraphics[width=0.6\textwidth]{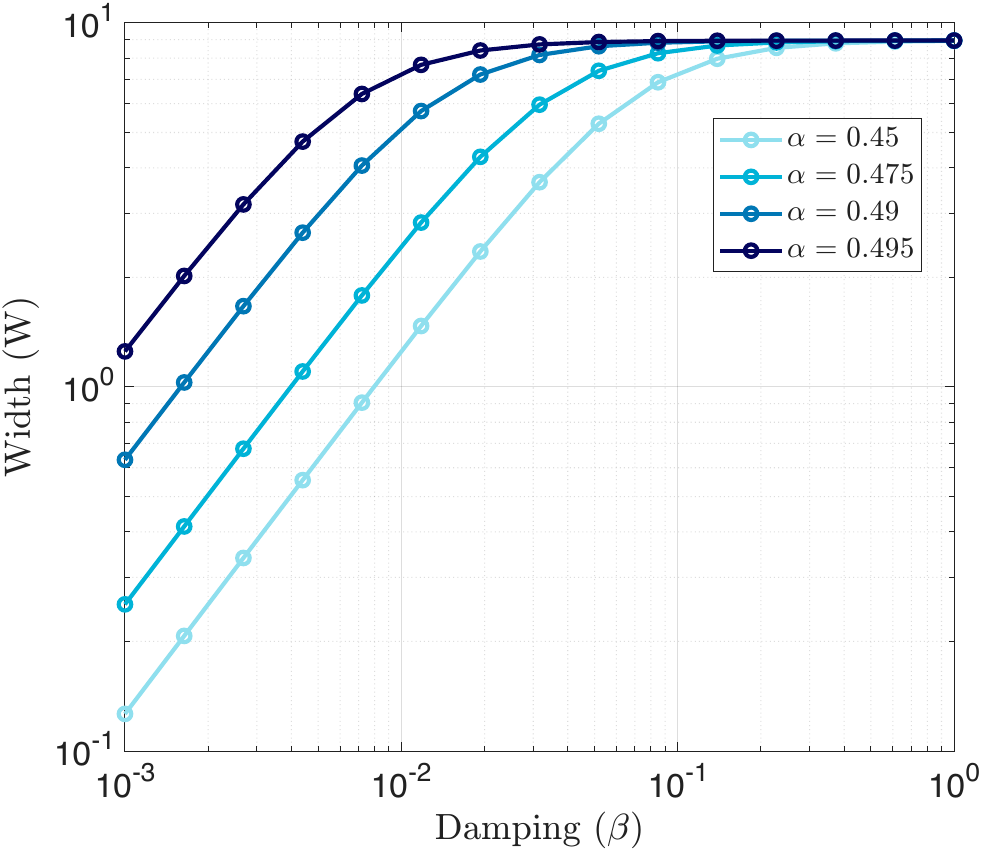}
\caption{\label{fig:SIWidth} Characteristic width $W$ of the propagating kink, analytical form from Eq. \ref{eq:SIW} compared to the full simulation (using Matlab's \textit{ode45}).}
\end{figure}
Fig. \ref{fig:SIfigEk} shows the kinetic energy trend for different $\alpha$ and $\beta$. Here $E_k$ diverges for a case without damping, where the kink approaches the propagation speed $c$. For a high damping case, instead, the kinetic energy drops with $v$ (as explained before), tending to 0. A different behavior can be seen for the elastic energy (Fig. \ref{fig:SIfigEe}), where it diverges for low damping (the width $W\rightarrow0$), while it reaches an asymptotic value when damping is high (the width converges to a finite value).

\clearpage
\newpage
\section{Lattice with defect: critical defect to pin a transition wave}
We consider the same system as before, but now we introduce a localized impurity at location $x_d$, with a potential $\epsilon$ times the potential of the baseline unit:
\begin{equation}
\ddot{u} + \beta \dot{u} - c^2 u_{,xx} + \left(1 + (\epsilon - 1)\delta(x - x_d)\right)\frac{dV(u)}{du} = 0,
\end{equation}
where $\delta$ is a regularized Dirac delta.
To calculate the threshold defect strength $\epsilon_{min}$ required to stop the kink soliton, we employ a blended approach that interpolates between two limiting regimes: the adiabatic (energy-balance) limit, applicable when damping effects during the kink-defect interaction are weak, and the force-balance limit, applicable when damping is strong enough to allow quasi-static adjustment of the kink profile. This blending is governed by a dimensionless parameter $\gamma$ that quantifies the effective damping over the interaction timescale. The dimensionless damping parameter is defined as: 
\begin{equation}
\gamma = \beta \frac{W}{v},
\end{equation}
representing the product of the damping coefficient $\beta$ and the interaction timescale $\tau = W / v$ (time for the kink to traverse its own width, approximating the defect interaction duration).

\subsection{Adiabatic Limit}
\begin{figure}
\centering
\includegraphics[width=0.6\textwidth]{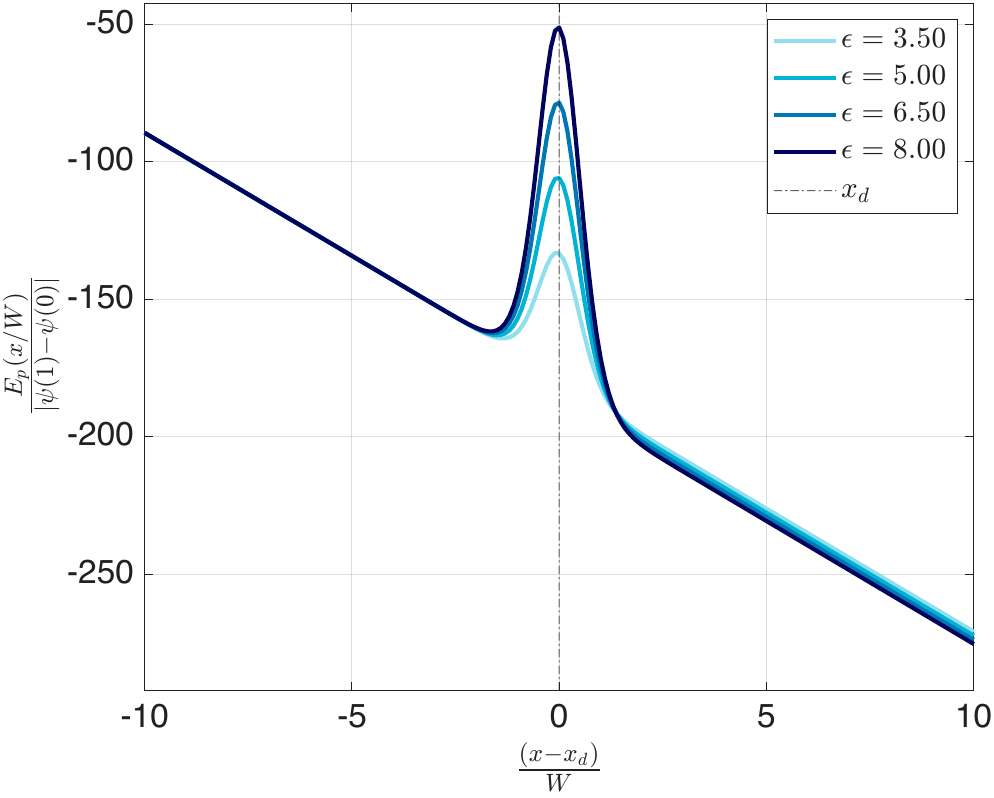}
\caption{\label{fig:SIEpdef} Potential energy $E_p$ of the kink interacting with the defect from Eq. \ref{eq:SIPotentialEnergydef}. $\alpha=0.495$.}
\end{figure}

In the low-damping regime ($\gamma \ll 1$) and still in a continuous regime, the kink interacts with the defect nearly conservatively. The threshold $\epsilon_{min}^{adiab}$ is determined by equating the kink's kinetic energy $E_k$ to the peak of the interaction potential $U_d(\xi_0)$ \cite{malomed1992perturbative} between the defect and the kink. $U_d(\xi_0)$ represents the effective interaction potential experienced by the kink (treated as a quasi-particle) due to the defect; this potential varies as a function of the distance between the kink's center of mass and the impurity.

\begin{equation}\label{eq:SIinteractionpotential}
U_d(\xi_0)=\int_{-\infty}^{\infty} (\epsilon_{min}^{adiab}-1)\delta(\xi-\xi_d)V(u(\xi-\xi_0))d\xi=(\epsilon_{min}^{adiab}-1)V(u(\xi_d-\xi_0)),
\end{equation}
with $\xi_d=\frac{x_d-vt}{W}$ the defect location and $\xi_0=\frac{x_0-vt_0}{W}$ the kink location. From the above equation, when $\xi_0\ll \xi_d$ (or $\xi_0\gg\xi_d$), $u(\xi_d-\xi_0)\approx0$ (or $1$), and thus $\frac{dV(u)}{du}\approx0$, indicating no interaction force between the kink and the defect. When $\xi_0$ is close to $\xi_d$ the defect exert a repulsive force on the kink.
The maximum of $U_d(\xi_0)$ is simply given by the maximum of the nonlinear potential $V(u)$. This happens when $u=\alpha$. The maximum of $U_d(\xi_0)$ is therefore: 
\[
max(U_d(z))=(\epsilon_{min}^{adiab}-1)V(u=\alpha)=(\epsilon_{min}^{adiab}-1)\frac{\alpha^3(2-\alpha)}{12}.
\]
\\
As calculated before, the kinetic energy is
\begin{equation}
E_k = \frac{v^2}{6 W}.
\end{equation}
Setting $E_k = max(U_d(\xi_0))$ yields
\begin{equation}
\epsilon_{min}^{adiab} = 1 + \frac{2 v^2}{W \alpha^3 (2 - \alpha)}.
\end{equation}
The potential energy in the lattice with a defect, shown in Fig. \ref{fig:SIEpdef}, is:
\begin{equation}\label{eq:SIPotentialEnergydef}
    E_p(\xi_0) = E_e(\xi_0) + \int_{-\infty}^{\infty} V(u(\xi-\xi_0))Wd\xi + U_d(\xi_0).
\end{equation}
The elastic energy part is constant, since the defect doesn't change the elastic coupling springs. The second term instead, can be easily calculated knowing that the difference in potential energy between a low and high state is $V(0)-V(1)=-\frac{2\alpha-1}{12}$. Therefore, the system potential energy will decrease linearly with the translation of the kink. 
 
The potential energy becomes:
\begin{equation}\label{eq:pot_en}
    E_p(\xi_0) = E_e  -\frac{2\alpha-1}{12}\xi_0 + U_d(\xi_0).
\end{equation}

\subsection{Force-Balance Limit}
In the high-damping regime ($\gamma \gg 1$), the kink adjusts quasi-statically to the defect, and the threshold $\epsilon_{min}^{force}$ is found by balancing the driving force from the potential asymmetry against the maximum opposing force from the defect.

The asymmetry driving force is:
\begin{equation}\label{eq:SIdrivingforce}
F_{driving}=\int_{-\infty}^{\infty} \frac{dV(u)}{du}\frac{du}{d\xi_0}Wd\xi_0=\frac{1-2\alpha }{12}
\end{equation}
which is positive for $\alpha < 0.5$ (driving the kink forward).

The force experienced by the kink from the defect can be calculated in a similar way, 
\begin{equation}\label{eq:SIdefectbarrierforce}
\begin{split}
    F_{barrier}(\xi_0)=\int_{-\infty}^{\infty} \delta(\xi-\xi_d)\frac{dV(u(\xi-\xi_0))}{du}\frac{du}{d\xi}\frac{d\xi}{W}=
\\
=\frac{dV(u(\xi_d-\xi_0))}{du}\frac{du(\xi_d-\xi_0)}{d\xi}\frac{1}{W}.
\end{split}
\end{equation}
What really interests us, is the maximum of the function $F_{barrier}$. If this is lower than the driving force $F_{driving}$ then the kink can slowly pass through the defect. The gradient $\frac{du}{d\xi} = -\frac{1}{2} \mbox{sech}^2\left(\xi_d - \xi_0\right)$ has its maximum at $\xi_0=\xi_d$, while the bistable force $\frac{dV(u)}{du}$ has its maximum for $u=\frac{1}{3}(\pm \sqrt{\alpha^2-\alpha+1}+\alpha+1)$ (we will use the sign $-$ because the kink moving from $0$ to $1$ will initially interact with the defect with smaller displacements). Therefore, the maximum of $F_{barrier}$ has to be calculated numerically. 

The maximum opposing force is $max(F_{barrier}(\xi_0))$ (positive peak, as it resists the negative driving).

When $F_{driving} + (\epsilon_{min}^{force} - 1) max(F_{barrier}(\xi_0))=0$ (balancing magnitudes),
\begin{equation}
\epsilon_{min}^{force} = 1 - \frac{F_{driving}}{max(F_{barrier}(\xi_0))}.
\end{equation}

\subsection{Blended Threshold}
To interpolate between regimes, we make use of an exponential blending of the form,
\begin{equation}
\epsilon_{min} = \epsilon_{min}^{adiab} e^{-\gamma} + \epsilon_{min}^{force} (1 - e^{-\gamma}),
\end{equation}
where $e^{-\gamma}$ is the weight for the adiabatic limit (high when $\gamma$ small), transitioning to force as $\gamma$ increases.

This approach captures the damping-dependent pinning, improving accuracy over pure energy balance, especially for intermediate $\beta$.

\begin{figure}
\centering
\includegraphics[width=0.6\textwidth]{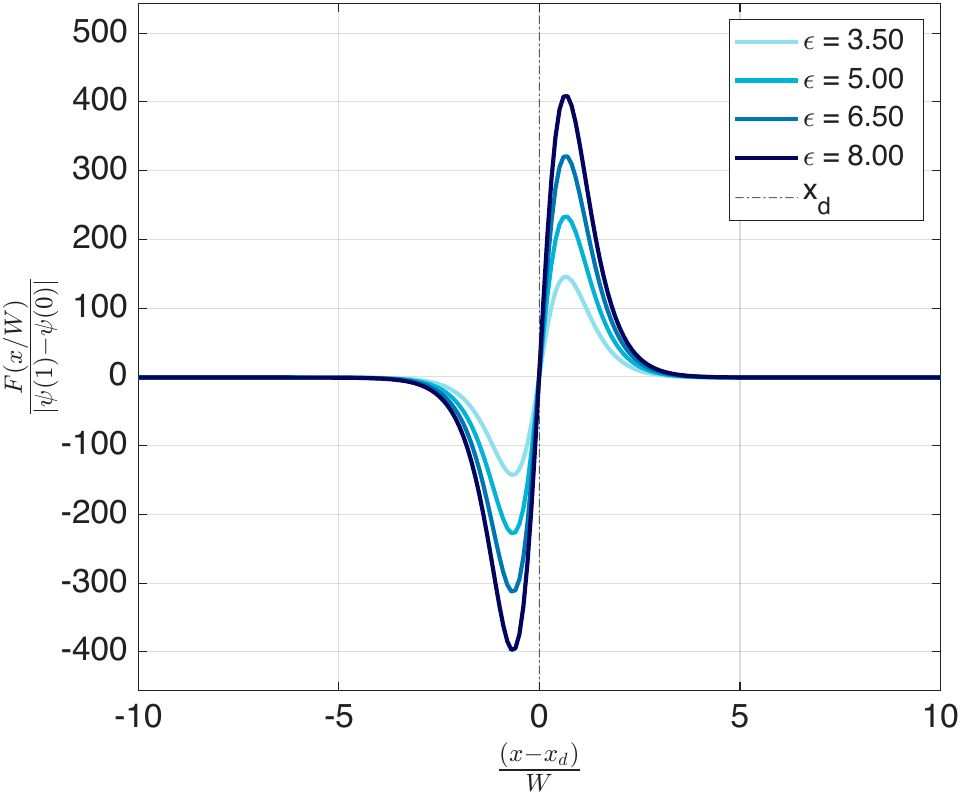}
\caption{\label{fig:SIFpdef} Interaction force $F_{barrier}$ of the kink interacting with the defect, from Eq. \ref{eq:SIdefectbarrierforce}. $\alpha=0.495$. }
\end{figure}
\begin{figure}
\centering
\includegraphics[width=0.6\textwidth]{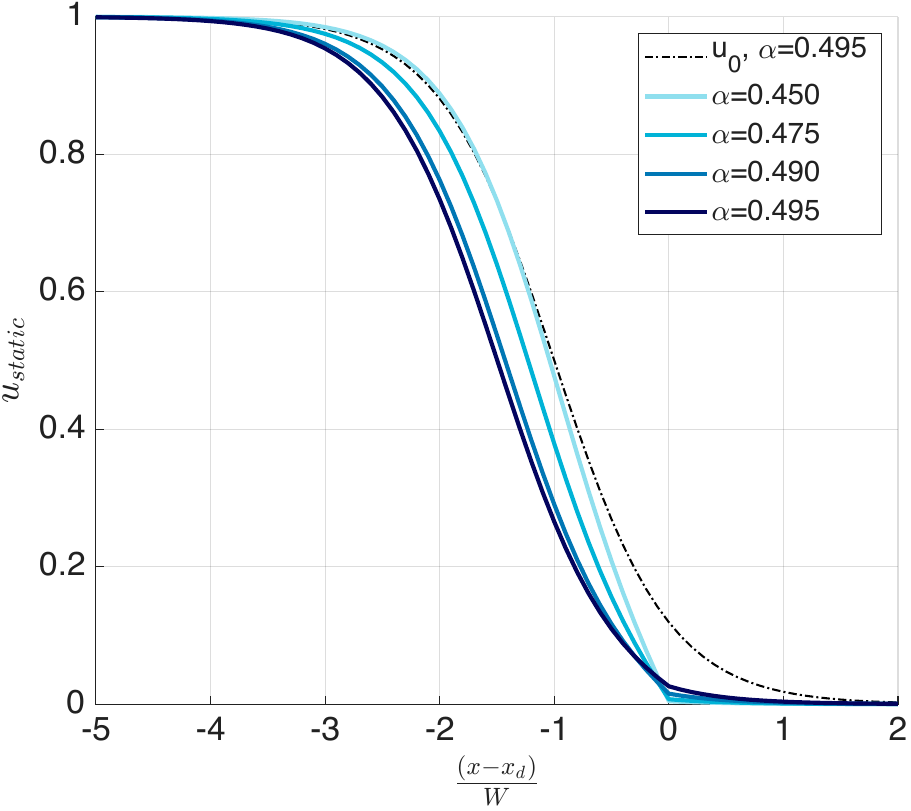}
\caption{\label{fig:SIu0alpha} Displacement profile of the pinned kinks on the defects $\epsilon=\epsilon_{min}$, for different values of $\alpha$. The dashed line correspond to the first guess of the numerical solver, which starts with a tentative shape equal to the analytical shape previously calculated for a system with no imperfections; the kink's center is $\epsilon_k=-W$. Setting the kink's center at the defect location doesn't not ensure the numerical convergence of the solver (the kink would be on an unstable equilibrium).}
\end{figure}

\clearpage
\newpage
\section{Perturbation analysis of a pinned kink}

\subsection{Shape of a pinned static kink}
We consider the previous system with the localized impurity at the location $x_d$ and with a potential $\epsilon$ times the potential of the baseline unit. When the kink is pinned by the defect, its deformed shape $u_0(x)$ can be calculated numerically from the static equation,

\begin{equation}
- c^2 u_{0,xx} + (1+(\epsilon-1)\delta(x-x_d))V'(u_0) = 0 .
\end{equation}
This is done using the MATLAB function $fsolve()$ with $u_k(x+W)$ as the initial guess, $u_k$ being the analytically, previously calculated, kink shape.

\subsection{Linear perturbation analysis}
Considering the full dynamic equation,
\begin{equation}\label{eq:SIeom_preperturbation}
\ddot{u} + \beta \dot{u}  - c^2 u_{,xx} + (1+(\epsilon-1)\delta(x-x_d))V'(u_0) = 0 ,
\end{equation}
we can study a perturbed solution $u^*=u_0(x)+\delta u(x,t)$.
Therefore the derivatives of this solution read
\begin{equation}
    \dot{u}^*=\frac{\partial u_0(x)}{\partial t} + \dot{\delta u}=\dot{\delta u} ,
\end{equation}
\begin{equation}
    \ddot{u}^*=\ddot{\delta u} ,
\end{equation}
\begin{equation}
    u^*_{,xx}=u_{0,xx} + \delta u_{,xx} .
\end{equation}
The nonlinear potential $V'(u^*)$ can be expanded as
\begin{equation}
\begin{split}
    V'(u^*)=V'(u_0+\delta u)=(u_0+\delta u)^3-(\alpha+1)(u_0+\delta u)^2+\alpha(u_0+\delta u) ,
    \\
    (u_0+\delta u)^3=u_0^3+3u_0^2\delta u+3u_0\delta u^2+\delta u^3 ,
    \\
    (u_0+\delta u)^2=u_0^2+2u_0\delta u+\delta u^2 .
\end{split} 
\end{equation}
Considering higher degrees as small:
\begin{equation}\label{eq:SIpotentialterms}
\begin{split}
    V'(u^*)=(u_0^3-(\alpha+1)u_0^2+\alpha u_0)+
    \\
    (3u_0^2-2(\alpha+1)u_0+\alpha)\delta u+(3u_0-(\alpha+1))\delta u^2+\delta u^3 = 
    \\
    = V'(u_0)+V''(u_0)\delta u+\cancel{\frac{V'''(u_0)}{2!}\delta u^2}+\cancel{\frac{V^{IV}(u_0)}{3!}\delta u^3} = V'(u_0)+V''(u_0)\delta u+\cancel{N(u_0,\delta u)}.
\end{split}
\end{equation}
Substituting all this in Eq. \ref{eq:SIeom_preperturbation}, and removing the equilibrium of the kink, gives:
\begin{equation}\label{eq:fulleq_perturbation}
    \ddot{\delta u}-c^2\delta u_{,xx}+\beta\dot{\delta u}+(1+(\epsilon-1)\delta(x-x_d))V''(u_0)\delta u=0 .
\end{equation}
Assuming solutions of the form $\delta u(x,t)=\phi(x)e^{\lambda t}$, the previous equation becomes:
\begin{equation} \label{eq:fulleq_perturbation2}
    \lambda^2 \phi-c^2\phi_{,xx}+\lambda\beta\phi+V_d''(u_0)\phi=0 ,
\end{equation}
where $V_d''(u_0) = (1+(\epsilon-1)\delta(x-x_d))V''(u_0)$. To numerically find the eigenmodes $\phi_i$ and eigenvalues $\lambda_i$, we discretize the system in N parts as done before for the discrete dispersion relation. The second order spatial derivative is,
\begin{equation}
\begin{split}
    \phi_{j,xx}=\frac{\phi_{j+1}-2\phi_j+\phi_{j-1}}{a_s^2} ,
\end{split}
\end{equation}
and once substituted in Eq.~\ref{eq:fulleq_perturbation2}, it becomes,
\begin{equation}\label{eq:sq_eigenproblem}
    (\lambda^2I+\beta \lambda I+A)\vec{\phi}=0 ,
\end{equation}
where $I$ is a $N\times N$ identity matrix, $A=-c^2D+diag(V_d''(u_0))$, and $D$ is the second derivative operator matrix.

We need to reduce the order of this eigenvalue problem to get to $M\vec{\phi}=\lambda N\vec{\phi}$:
\begin{equation}
    \phi*=
    \begin{bmatrix}
    \phi\\
    \lambda \phi
    \end{bmatrix}
    =
    \begin{bmatrix}
    \phi_1\\
    \phi_2
    \end{bmatrix}.
\end{equation}
Eq. \ref{eq:sq_eigenproblem} becomes:
\begin{equation}
    \begin{cases}
    \begin{aligned}
        \phi_2=\lambda \phi_1 \\
        \lambda I\phi_2+\beta I\phi_2+A\phi_1=0
    \end{aligned},
    \end{cases}
\end{equation}
\begin{equation}
    \begin{bmatrix}
    0 & I\\
    -A & -\beta I
    \end{bmatrix}
    \begin{bmatrix}
    \phi_1\\
    \phi_2
    \end{bmatrix}=\lambda
    \begin{bmatrix}
    I & 0\\
    0 & I
    \end{bmatrix}
    \begin{bmatrix}
    \phi_1\\
    \phi_2
    \end{bmatrix}.
\end{equation}
The numerically derived eigenvalues and eigenvectors will be called from now on, respectively, $\lambda_j$ and $\phi_j(x)$. The eigenvectors and values resulting from this calculation are of three types: a finite-frequency Goldstone translational mode $\phi_T$, an internal mode $\phi_I$, and then de-localized (phonon) modes. the characteristic shape of these modes, and their effect on the kink's shape are illustrated in Fig. \ref{fig:fig2}d,e.

\newpage
\section{Scattering Analysis}

We want to study how phonons can excite the pinned kink and depin it. To do so we need to find the traveling wave form of the phonons. We perturb around the static kink: $u(x, t) =
u_0(x) + \Re(\phi_P (x)e^{-i\omega t})$, where $\phi_P (x)$ is small complex amplitude. Substituting into the full PDE (ignoring damping ($\beta = 0$) as damping would add small imaginary frequency shift because of the weak damping considered):
\begin{equation}
-\omega^2\phi_P - c^2\phi_{P,xx} + (1 + (\epsilon - 1)\delta(x - x_d))V''(u_0)\phi_P = 0,
\end{equation}
The phonon equation is:
\begin{equation}
\phi_{P,xx} + \frac{\omega^2 - V''_d (u_0)}{c^2} \phi_P = 0.
\end{equation}

\subsection{Defining the Envelope $\Phi (x)$}

We define the envelope $\Phi(x)$ by factoring the right-going carrier wave (incident from the left):
\begin{equation}
\phi_P (x) = \Phi(x) e^{i q_L x}.
\end{equation}
To analyze the scattering of phonons by the pinned kink in the discrete lattice, we employ a transfer matrix method. This approach allows us to compute the reflection and transmission coefficients for incident phonons, as well as the spatial envelope of the perturbation field.

The potential $V_{dn} = V_d(n)$ is sampled at each of the discrete sites. The discrete perturbation amplitude is $\phi_n$, such that the full perturbation is $u_n(t) = u_{0n} + \Re(\phi_n e^{-i \omega_P t})$. The discrete equation for $\phi_n$ is:
\begin{equation}
\omega_P^2 \phi_n = - c^2 (\phi_{n+1} - 2\phi_n + \phi_{n-1}) + V_{dn}'' \phi_n.
\end{equation}
Where $ V_{dn}''= V_{d}''(u_0(x=na_s))$. Rearranging:
\begin{equation}
\phi_{n+1} = \left(2 - \frac{\omega_P^2 - V_{dn}''}{c^2}\right) \phi_n - \phi_{n-1}.
\end{equation}
This can be written in matrix form as
\begin{equation}
\begin{pmatrix}
\phi_{n+1} \\
\phi_n
\end{pmatrix}
= M_n
\begin{pmatrix}
\phi_n \\
\phi_{n-1}
\end{pmatrix},
\end{equation}
where
\begin{equation}
M_n = \begin{pmatrix}
m_n & -1 \\
1 & 0
\end{pmatrix}, \quad m_n = 2 - \frac{\omega_P^2 - V_{dn}''}{c^2}.
\end{equation}
The total transfer matrix $M_\text{total}$ is the product $M_L \cdots M_{-L}$. We consider phonons incident from the left ($x \to -\infty$), where the asymptotic behavior is wavelike with wavenumber $q_L$ satisfying
\begin{equation}
q_L = \frac{1}{a_s}\arccos \Big(1 - \frac{\omega_P^2 - (1 - \alpha)}{2 c^2}\Big).
\end{equation}
Similarly for the right side with $q_R$ and $1-\alpha$ replaced by $\alpha$.

The incident wave at the left boundary ($n=-L$) is represented by vectors for incident, reflected, and transmitted parts. We solve the system for the full field:
\begin{equation}
M_\text{total} \begin{pmatrix} e^{i q_L (-L)} \\ e^{i q_L (-L - 1)} \end{pmatrix} + M_\text{total}R \begin{pmatrix} e^{-i q_L (-L)} \\ e^{-i q_L (-L - 1)} \end{pmatrix} = T \begin{pmatrix} e^{i q_R (L + 1)} \\ e^{i q_R L} \end{pmatrix} ,
\end{equation}
yielding the complex transmission $T$ and reflection $R$ coefficients.

Now we substitute $\phi_k = \Phi_k e^{i q_L k}$ for all indices $k = n+1, n, n-1$:
\begin{equation}
\Phi_{n+1} e^{i q_L (n+1)} = m_n \Phi_n e^{i q_L n} - \Phi_{n-1} e^{i q_L (n-1)},
\end{equation}
and divide both sides by the common factor $e^{i q_L (n+1)}$ to isolate terms involving $\Phi$:
\begin{equation}
\Phi_{n+1} = m_n \Phi_n \frac{e^{i q_L n}}{e^{i q_L (n+1)}} - \Phi_{n-1} \frac{e^{i q_L (n-1)}}{e^{i q_L (n+1)}}.
\end{equation}
The equation simplifies to:
\begin{equation}
\Phi_{n+1} = m_n \Phi_n e^{-i \theta_L} - \Phi_{n-1} e^{-2 i \theta_L},
\end{equation}
giving:
\begin{equation}
\begin{pmatrix}
\Phi_{n+1} \\
\Phi_n
\end{pmatrix}
= \begin{pmatrix}
e^{-i \theta_L} m_n & -e^{-2 i \theta_L} \\
1 & 0
\end{pmatrix}
\begin{pmatrix}
\Phi_n \\
\Phi_{n-1}
\end{pmatrix}.
\end{equation}

\subsection{Damping correction}
These phonons may travel a long distance from the boundary to the kink, and are therefore attenuated by damping.
We can re-write Eq. \ref{eq:linearwaves} by adding to the initial PDE the damping term: \( u_j(t) = \Re \left[ A e^{i (\tilde{q} j \, a_s - \omega t)} \right] \), with complex \( \tilde{q} = q + i \gamma \).
The equation becomes:
\begin{equation}
    -\omega^2 - i \beta \omega + \frac{2 c^2}{a_s^2} (1 - \cos(\tilde{q_L}a_s)) + (1-\alpha) = 0.
\end{equation}
For small $\beta$, $\cos(\tilde{q_L}a_s) = \cos(q_La_s + i \gamma_L a_s) \approx \cos(q_La_s) - i \sin(q_La_s) \gamma_L a_s $. The third term becomes \(\frac{2 c^2}{a_s^2} (1 - \cos(q_La_s)) + i \frac{2 c^2}{a_s^2} \gamma_L a_s \sin(q_La_s)\). The imaginary part of the equation is:
\begin{equation}
    -\beta \omega + \frac{2 c^2}{a_S^2} \gamma_L a_s \sin(q_La_s) = 0,
\end{equation}
yielding \(\gamma_L = \frac{\beta \omega \, a_s}{2 c^2 \sin(q_La_s)}\) and \(\gamma_R = \frac{\beta \omega \, a_s}{2 c^2 \sin(q_R \, a_s)}\). 

Utilizing the continuum approximation, instead:
\begin{equation}
    q_{L\beta}^2 = \frac{\omega_P^2 + i \omega_P \beta - (1 - \alpha)}{c^2 }=q_{L}+\frac{i \omega_P \beta}{c^2}.
\end{equation}
For small $\beta$, we can approximate $q_{L\beta} \approx q_L+i\frac{\omega_P \beta}{2c^2q_L}= q_L+i\gamma_L$. The amplitudes are therefore reduced by a coefficient $e^{\frac{- \omega_P \beta}{2c^2q_L}x}$. Because the kink is pinned next to the defect, the decrease in amplitudes will be $\approx e^{\frac{- \omega_P \beta}{2c^2q_L}x_d}$.
On the left side ($x < x_d$), we separate the incident and reflected components of the envelope $\Phi$: the incident part (propagating rightward) is multiplied by $e^{-\gamma_L x}$, and the reflected part (decaying leftward) by $e^{\gamma_L x}$. On the right side ($x \geq 0$), the transmitted wave is multiplied by $e^{-\gamma_R x}$ (decaying rightward). The damped envelope $\Phi^\text{damped}$ is normalized such that $|\Phi^\text{damped}_{-L}| = 1$ at the left boundary.

\begin{figure}
\includegraphics[width=0.6\textwidth]{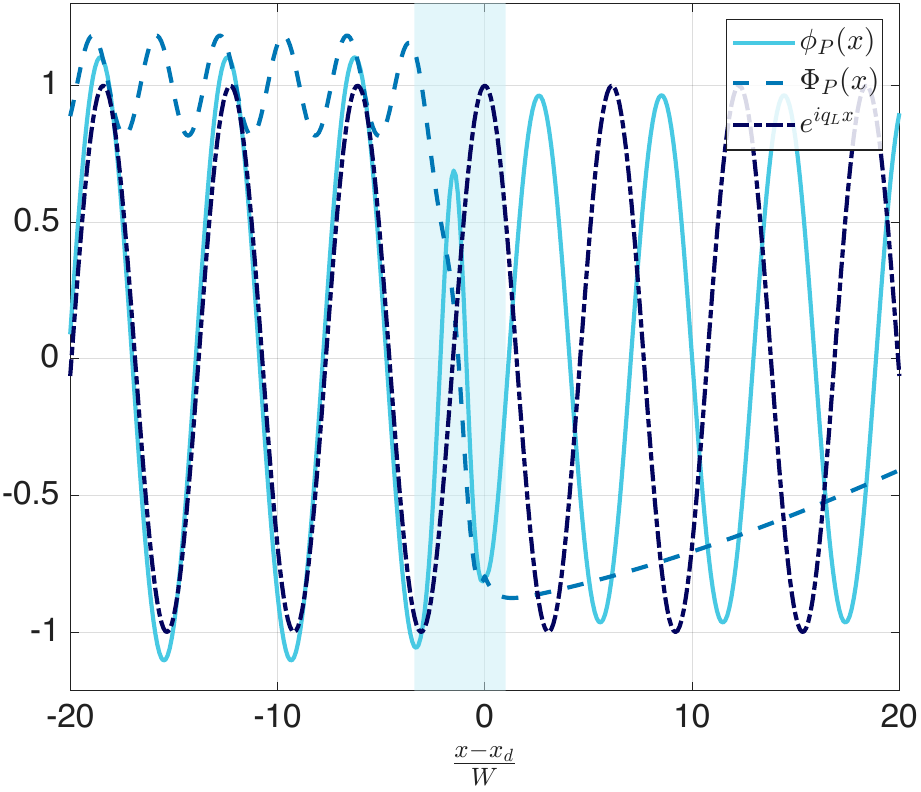}
\centering
\caption{\label{fig:SIscatphi} Phonon modulation in a lattice with a pinned kink and a defect. The light blue shade indicates where the kink is ($0.01<u<0.99$)}
\end{figure}
\begin{figure}
\centering
\includegraphics[width=0.6\textwidth]{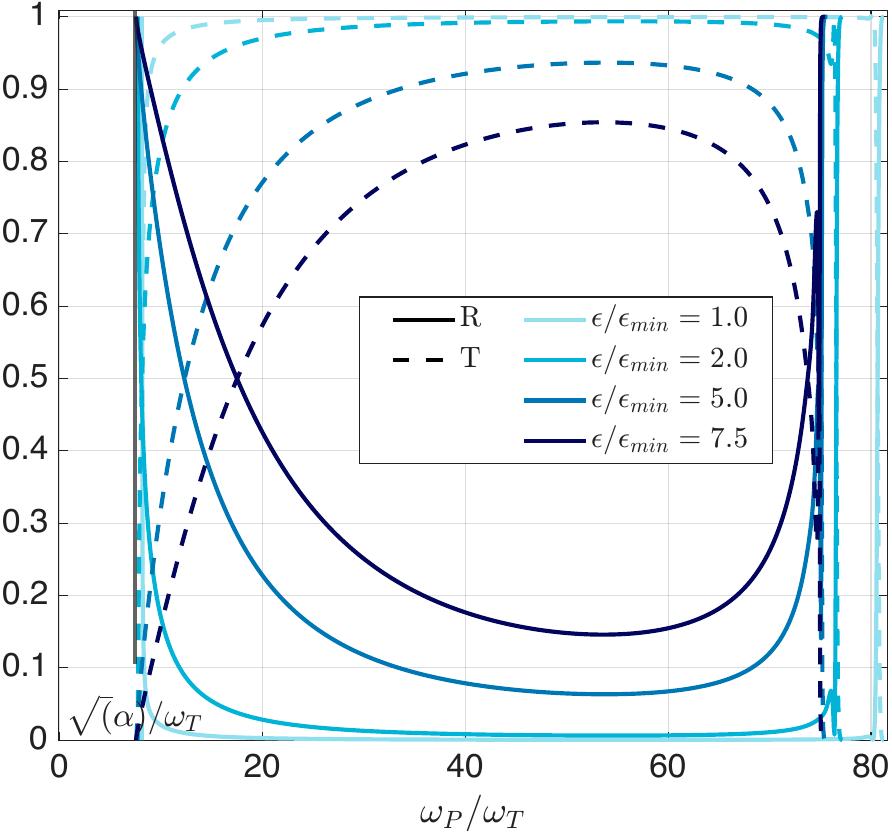}
\caption{\label{fig:SIRT} Reflection $R$ and transmission $T$ coefficients for different defect strength.}
\end{figure}
\begin{figure}
\centering
\includegraphics[width=0.6\textwidth]{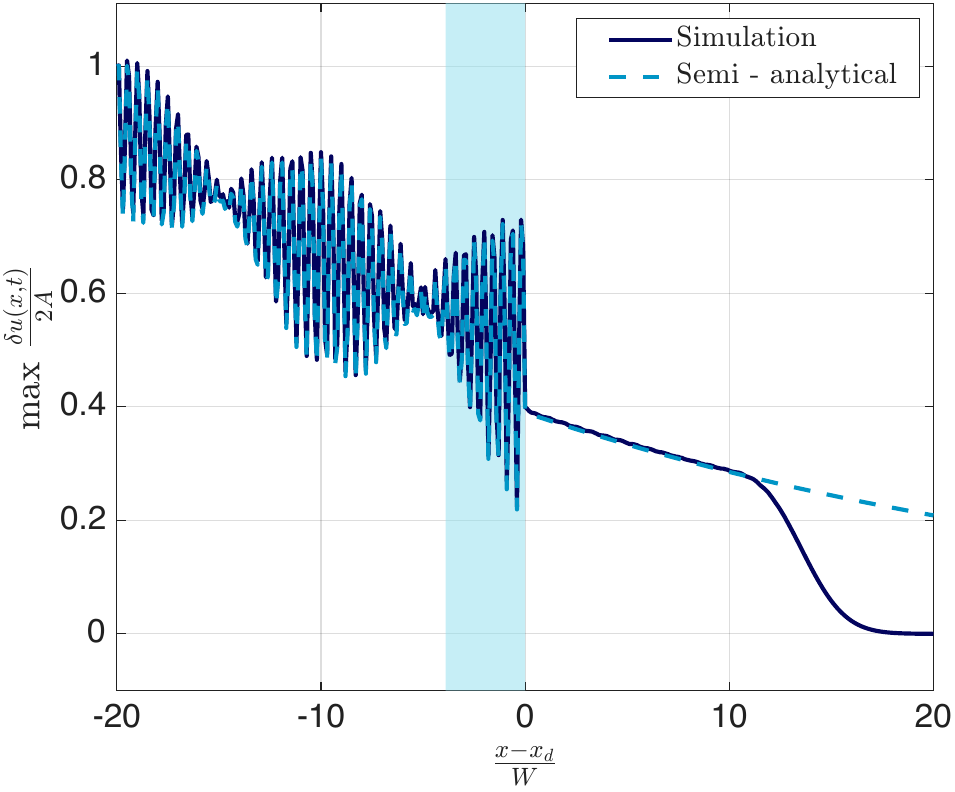}
\caption{\label{fig:SIPhononVal} Validation of the beating phonons interacting with the kink and defect with damping correction. Here the perturbation $\delta u(x,t)$ is introduced in the simulation and run over time, then the maximum amplitude for each unit is plotted and compared with the analytical form. Here $A=${5e-4} and $\epsilon=7.5\epsilon_{min}$ is used. The difference in decay for $\frac{x-x_d}{W}>10$ is caused by a sponge layer used in the full simulation to avoid reflected waves from the boundary.}
\end{figure}

\clearpage
\newpage
\section{Reduced order model for the T mode}
\subsection{The effect of phonon on T mode}
To study the dynamics of the pinned kink, we focus on the translational-like (T) mode because it encapsulates the collective displacement of the kink as a whole, effectively governing its motion across the defect. 
From Eq. \ref{eq:fulleq_perturbation}, we derive,
\begin{equation}
    [-c^2\partial{,xx}+(1+(\epsilon-1)\delta(x-x_d))V''(u_0)]\phi_j=-\lambda_j^2 \phi_j-\lambda_j\beta\phi_j,
\end{equation}
where the eigenvalues $\lambda_j$ are related to the natural undamped frequency $\omega_j$ of each mode as,
\begin{equation} \label{eq:lambda_j_roots}
    \lambda_j = -\frac{\beta}{2} \pm \sqrt{\left(\frac{\beta}{2}\right)^2 - \omega_j^2} ,
\end{equation}
with $Re(\lambda_j)$ the decay rate and $Im(\lambda_j) $ the actual oscillation frequency of the damped mode.

We expand the perturbation $\delta u(x,t)$ using these spatial eigenvectors $\phi(x)$ found before from the stability and scattering analysis:
\begin{equation} \label{eq:eta_expansion}
    \delta u(x,t) =\sum_J \delta u_J(x,t) = a_T(t) \phi_T(x)+a_I(t) \phi_I(x) + \sum_j Re(a_{Pj} \Phi_{Pj}(x)e^{i(q_{Lj}x-\omega_jt)}) ,
\end{equation}
where $J=\{T, I, P1, P2, ...\}$. The $\phi_J(x)$ are assumed to form a complete orthonormal basis such that $\int \phi_j(x) \phi_k(x) dx = \delta_{jk}$ (after normalization). If we substitute Eq. \eqref{eq:eta_expansion} into the linearized PDE Eq. \eqref{eq:fulleq_perturbation}, multiply by $\phi_T(x)$, and integrate over $x$ (using orthogonality) the equation yields a set of ordinary differential equations (ODE) for the different modes in the perturbation decomposition. Focusing only on the translational mode $a_T(t)$,
\begin{equation} \label{eq:a_j_ode_linear}
    \ddot{a}_{T}(t) + \beta \dot{a}_{T}(t) + (-\lambda_T^2-\lambda_T\beta) a_T(t) =\ddot{a}_{T}(t) + \beta \dot{a}_{T}(t) + \omega_T^2 a_T(t) = 0 ,
\end{equation}
which is the ODE for free vibration of a linear oscillator with linear damping. Since we hope to investigate the coupling/effect of the phonon modes with the translational ($T$) mode, we have to consider the non-linear terms in $N(u_0, \delta u)$ (the previously discarded term in Eq. \ref{eq:SIpotentialterms}). Here we only consider the first term which is of  order 2,
\begin{equation}
    N(u_0, \delta u) = \frac{V'''(u_0)}{2}\delta u^2+O(\epsilon^3) \approx \frac{V'''(u_0)}{2} \big( \sum_J \delta u_J(x,t) \big)^2 ,
\end{equation}
projecting them onto the translational mode, finding a non-linear forcing term $F_T(t)$.
\begin{equation} \label{eq:a_k_ode_nonlinear_full}
    \ddot{a}_{T}(t) + \beta \dot{a}_{T}(t) + \omega_T^2 a_T(t) = F_T(t),
\end{equation}
where $F_T(t)$ is obtained by the projection as:
\begin{equation} \label{eq:Nk_definition1}
    F_T(t) = - \int_{-\infty}^{\infty} \phi_T(x) N\big(u_0(x), \sum_J \delta u_{PJ}(x,t)\big) dx.
\end{equation}
As it can be seen from this equation, because the forcing term is $O(\epsilon^2)$, $a_T(t)$ is also $O(\epsilon^2)$; the higher order terms ($\propto a_Pa_T$ and $\propto a_T^2$) that arise form the squared perturbation are $O(\epsilon^3)$ and $O(\epsilon^4)$ respectively. This is also true for the terms $\propto a_Pa_I$ and $\propto a_I^2$. Therefore, $F_T(t)$, in the approximation $O(\epsilon^2)$, is only dependent on the phonons, yielding therefore an equation of motion that is still a linear ODE.

It should also be noted that, although we have considered the 2nd-order nonlinear term of the perturbed potential $\delta V$, the equation of motion for the T-mode Eq.~(\ref{eq:a_k_ode_nonlinear_full}) is still a linear ODE because $F_T$ doesn't depend on $a_T$ anymore.

\subsection{Excitation of the T mode using beats}
From Section 2 of this document, we can find $q_{Lj}$ and $q_{Rj}$ (wavevectors of the left side ($u=1$) and of the right side ($u=0$)) and their relations to the frequencies $\omega_j$.
The $T$ mode eigenfrequency is found to be much lower compared to the minimum allowed frequency $\omega_{min}=min(\sqrt{1-\alpha},\sqrt{\alpha})$ from the dispersion relation. Therefore, to excite this low frequency mode using phonons we need to modulate allowed phonons that can travel through the system with a low frequency modulation that matches the $T$ mode frequency.
This can be achieved directly by using two beating phonons with $\Delta \omega = |\omega_{j_1} - \omega_{j_2}| \approx \omega_T$.
Each phonon mode is decomposed as:
\begin{equation}
\begin{split}
            \delta u_{Pk}(x, t) = \mathfrak{Re}(a_{Pk} \Phi_{Pk}(x)e^{i(q_{Lk}x-\omega_kt)}) = 
    \\
            a_{Pk} [G_k(x) \cos \theta_k - U_k(x) \sin \theta_k],
\end{split}
\end{equation}
where $k=1,2$ refer to the phonons $j_1$ and $j_2$. $\theta_k = q_{Lk} x - \omega_k t$, $G_k(x) = \mathfrak{Re}(\Phi(x))$ is the cosine-modulated envelope, and $U_k(x) = \mathfrak{Im}(\Phi(x))$ is the sine-modulated.
\\
The total perturbation is:
\begin{equation}\label{eq:total ptb}
\delta u_B(x, t) = a_{P1} [G_1 \cos \theta_1 - U_1 \sin \theta_1] + a_{P2} [G_2 \cos \theta_2 - U_2 \sin \theta_2].
\end{equation}
\\
To compute $F_T(t)$, we need to calculate $\delta u_B^2$, and expand it into self and cross terms.
The self term, for each $k=1,2$:
$$[G_k \cos \theta_k - U_k \sin \theta_k]^2 = G_k^2 \cos^2 \theta_k + U_k^2 \sin^2 \theta_k - 2 G_k U_k \cos \theta_k \sin \theta_k.$$
Using trigonometric identities we find:
$$\frac{1}{2} (G_k^2 + U_k^2) + \frac{1}{2} (G_k^2 - U_k^2) \cos 2\theta_k - G_k U_k \sin 2\theta_k.$$
We can separate the time dependence from the space dependence:
$$\cos 2\theta_k = \cos(2 q_{Lk} x) \cos(2 \omega_k t) + \sin(2 q_{Lk} x) \sin(2 \omega_k t),$$
$$\sin 2\theta_k = \sin(2 q_{Lk} x) \cos(2 \omega_k t) - \cos(2 q_{Lk} x) \sin(2 \omega_k t).$$
The coefficient of $\cos(2 \omega_k t)$ in the self expansion is $\frac{1}{2} (G_k^2 - U_k^2) \cos(2 q_{Lk} x) - G_k U_k \sin(2 q_{Lk} x)$,
and of $\sin(2 \omega_k t)$ is $\frac{1}{2} (G_k^2 - U_k^2) \sin(2 q_{Lk} x) + G_k U_k \cos(2 q_{Lk} x)$. Thus,
\begin{equation}\label{eq:FT_self}
F_T^{\text{self},k}(t) = a_{Pk}^2 \left[ K_{\text{stat}}^{kk} + K_{\text{cos}}^{kk} \cos(2 \omega_k t) + K_{\text{sin}}^{kk} \sin(2 \omega_k t) \right],
\end{equation}
where
$$K_{\text{stat}}^{kk} = - \int \frac{1}{2}\phi_T V_d''' \frac{1}{2}(G_k^2 + U_k^2) \, dx,$$
$$K_{\text{cos}}^{kk} = - \int \frac{1}{2}\phi_T V_d''' \left[ \frac{1}{2} (G_k^2 - U_k^2) \cos(2 q_{Lk} x) - G_k U_k \sin(2 q_{Lk} x) \right] dx,$$
$$K_{\text{sin}}^{kk} = - \int \frac{1}{2}\phi_T V_d''' \left[ \frac{1}{2} (G_k^2 - U_k^2) \sin(2 q_{Lk} x) + G_k U_k \cos(2 q_{Lk} x) \right] dx.$$
\\
The cross term is:
$$2 a_{P1} a_{P2} [G_1 \cos \theta_1 - U_1 \sin \theta_1][G_2 \cos \theta_2 - U_2 \sin \theta_2].$$
Expanding products using Werner identities gives:
\[
    \begin{split}
        a_{P1} a_{P2} \Big[ (G_1 G_2 - U_1 U_2) \cos(\theta_1 + \theta_2) + (-G_1 U_2 - U_1 G_2) \sin(\theta_1 + \theta_2) +
        \\
        + (G_1 G_2 + U_1 U_2) \cos(\theta_1 - \theta_2) + (G_1 U_2 - U_1 G_2) \sin(\theta_1 - \theta_2) \Big].
    \end{split}
\]
Here, $\theta_1 + \theta_2 = (q_{L1} + q_{L2}) x - (\omega_1 + \omega_2) t$ is a high frequency term, $\theta_1 - \theta_2 = \Delta q_L x - \Delta \omega t$ a low frequency, with $\Delta q_L = q_{L1} - q_{L2}$, $\Delta \omega = \omega_1 - \omega_2 > 0$.
Expanding:
$$\cos(\theta_1 \pm \theta_2) = \cos( (q_{L1} \pm q_{L2}) x ) \cos( (\omega_1 \pm \omega_2) t ) + \sin( (q_{L1} \pm q_{L2}) x ) \sin( (\omega_1 \pm \omega_2) t ),$$
$$\sin(\theta_1 \pm \theta_2) = \sin( (q_{L1} \pm q_{L2}) x ) \cos( (\omega_1 \pm \omega_2) t ) - \cos( (q_{L1} \pm q_{L2}) x ) \sin( (\omega_1 \pm \omega_2) t ).$$
The cross contribution is:
\begin{equation}\label{eq:FT_cross}
F_T^{\text{cross}}(t) = a_{P1} a_{P2} \Big[ K_{\text{cos},+}^{12} \cos((\omega_1 + \omega_2) t) + K_{\text{sin},+}^{12} \sin((\omega_1 + \omega_2) t) + K_{\text{cos},-}^{12} \cos(\Delta \omega t) + K_{\text{sin},-}^{12} \sin(\Delta \omega t) \Big],
\end{equation}
where the coefficients for the high frequency terms are:
$$K_{\text{cos},+}^{12} = -\int \frac{1}{2} \phi_T V_d''' (G_1 G_2 - U_1 U_2) \cos( (q_{L1} + q_{L2}) x ) \, dx - \int \frac{1}{2}\phi_T V_d''' (-G_1 U_2 - U_1 G_2) \sin( (q_{L1} + q_{L2}) x ) \, dx,$$
$$K_{\text{sin},+}^{12} = -\int \frac{1}{2}\phi_T V_d''' (G_1 G_2 - U_1 U_2) \sin( (q_{L1} + q_{L2}) x ) \, dx + \int \frac{1}{2}\phi_T V_d''' (-G_1 U_2 - U_1 G_2) \cos( (q_{L1} + q_{L2}) x ) \, dx,$$
and for the low frequency term: 
$$K_{\text{cos},-}^{12} = -\int \frac{1}{2}\phi_T V_d''' (G_1 G_2 + U_1 U_2) \cos( \Delta q_L x ) \, dx - \int \frac{1}{2}\phi_T V_d''' (G_1 U_2 - U_1 G_2) \sin( \Delta q_L x ) \, dx,$$
$$K_{\text{sin},-}^{12} = -\int \frac{1}{2}\phi_T V_d''' (G_1 G_2 + U_1 U_2) \sin( \Delta q_L x ) \, dx + \int \frac{1}{2} \phi_T V_d''' (G_1 U_2 - U_1 G_2) \cos( \Delta q_L x ) \, dx.$$
Substituting the forcing back into the equation of motion, and dividing it into three terms (static forcing, non-resonant (high-frequency) forcing, resonant forcing (beating frequency $\Delta \omega$)), we obtain:
\\
\begin{flalign} \label{eq:aP_ode_forced}
            &\ddot{a}_T(t) + \beta \dot{a}_T(t) + \omega_T^2 a_T(t) =F_{Tstat}+F_{Tres}(t)+F_{Tnon-res}(t)=&&
            \\
            =& \Big[  a_{P1}^2 K_{\text{stat}}^{11} + a_{P2}^2 K_{\text{stat}}^{22}\Big] + &&
            \\
            & + \Big[ a_{P1}^2K_{\text{cos}}^{11} \cos(2 \omega_1 t) + a_{P1}^2K_{\text{sin}}^{11} \sin(2 \omega_1 t) + a_{P2}^2K_{\text{cos}}^{22} \cos(2 \omega_2 t) + a_{P2}^2K_{\text{sin}}^{22} \sin(2 \omega_2 t) + &&
            \\
            &+ a_{P1} a_{P2} K_{\text{cos},+}^{12} \cos((\omega_1 + \omega_2) t) + a_{P1} a_{P2} K_{\text{sin},+}^{12} \sin((\omega_1 + \omega_2) t) \Big] +&&
            \\
            &+ a_{P1} a_{P2} \Big[ K_{\text{cos},-}^{12} \cos(\Delta \omega t) + K_{\text{sin},-}^{12} \sin(\Delta \omega t) \Big].&&
\end{flalign}
The above ODE describes the forced oscillations of a linear oscillator with linear damping. The steady state amplitude of the system can be calculated using $X(\omega_F)=\frac{F_0}{\sqrt{(\omega_n^2-\omega_F^2)^2+(\beta \omega_F)^2}}$, where $X$ is the response amplitude, $\omega_F$ is the forcing frequency, and $\omega_n$ is the natural frequency of the system. To achieve the maximum response of the system, we need the forcing frequency (here, the beating frequency $\Delta \omega$) to match the system natural frequency (the translational mode frequency $\omega_T$). Therefore, the selection of the pair of the two beating phonons ($1$ and $2$) is determined by finding all the pairs which lead to a beat frequency $\Delta \omega = \omega_T$ and, from these, the pair that maximizes $(K_{\text{cos},-}^{12})^2 + (K_{\text{sin},-}^{12})^2$.

The steady-state solution is:
\begin{equation}\label{eq:SIamplROM}
    a_T(t) \approx \frac{a_{P1}^2 K_{\text{stat}}^{11} + a_{P2}^2 K_{\text{stat}}^{22}}{\omega_T^2} + \frac{ a_{P1} a_{P2} \sqrt{ (K_{\text{cos},-}^{12})^2 + (K_{\text{sin},-}^{12})^2 } }{\beta \omega_T } \cos(\omega_T t + \phi + \psi),
\end{equation}
where $\phi = \tan^{-1}(K_{\text{sin},-}^{12} / K_{\text{cos},-}^{12})$, $\psi = \tan^{-1} \left( \frac{\beta \omega_T}{\omega_T^2 - \omega_T^2} \right)=\tan^{-1}(\infty)=\frac{\pi}{2}$.
The high-frequency component is not considered because $(\omega_i+\omega_j)^2 \gg \omega_T^2$.

\begin{figure}
\centering
\includegraphics[width=0.6\textwidth]{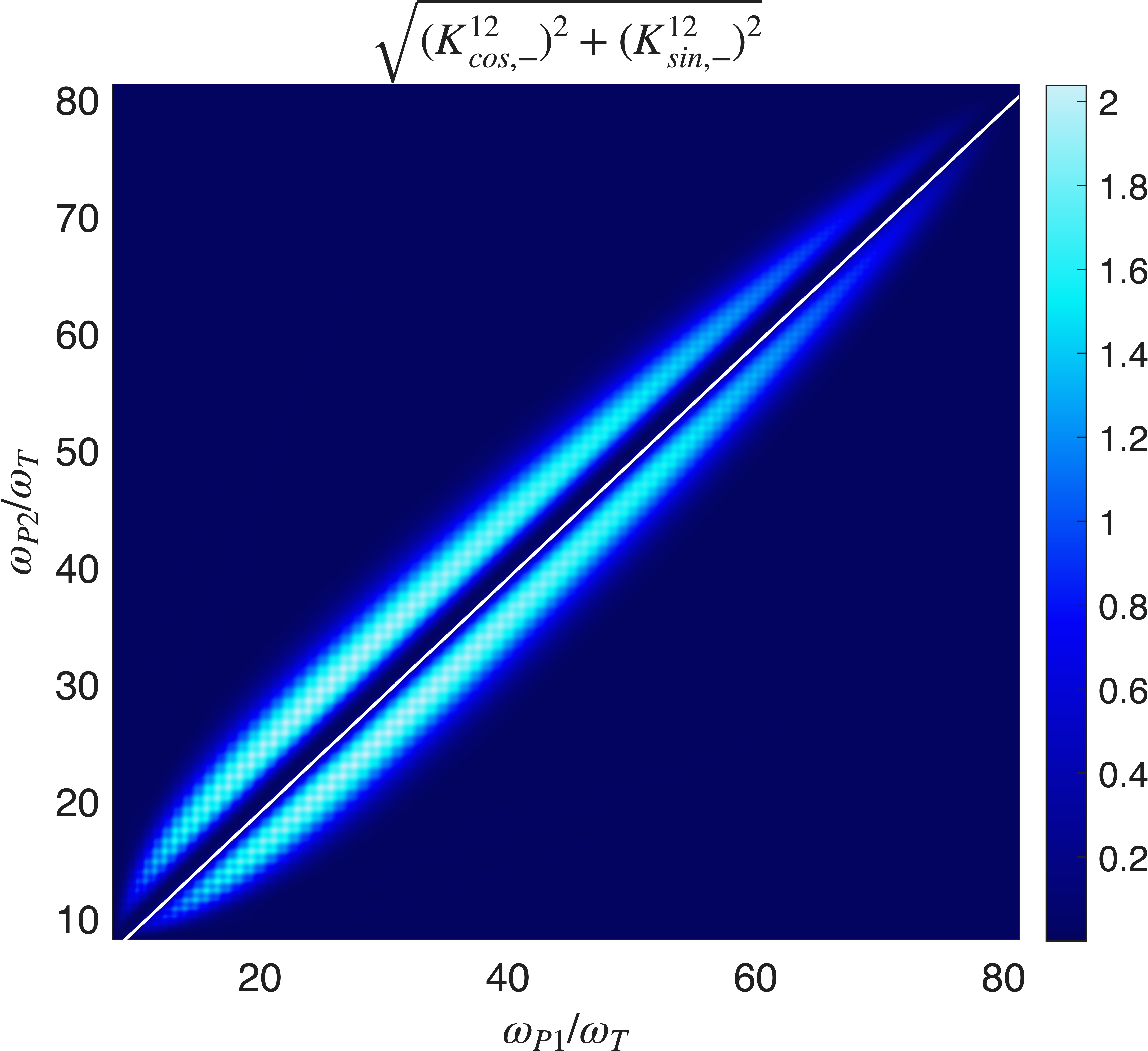}
\caption{\label{fig:SIKdyn} How to select the beating phonon pair. All the possible pairs couple with the kink, the color bar shows the coupling strength. To determine the optimal pair we fix a beating frequency (the white line for example is $\Delta \omega = \omega_T$), and on this line we find the pair that maximizes the coefficient $\sqrt{(K_{\text{cos},-}^{12})^2 + (K_{\text{sin},-}^{12})^2}$. For this plot we used $\epsilon=\epsilon_{min}$, $\alpha=0.495$, $\beta=0.02$.}
\end{figure}
\begin{figure}
\centering
\includegraphics[width=0.6\textwidth]{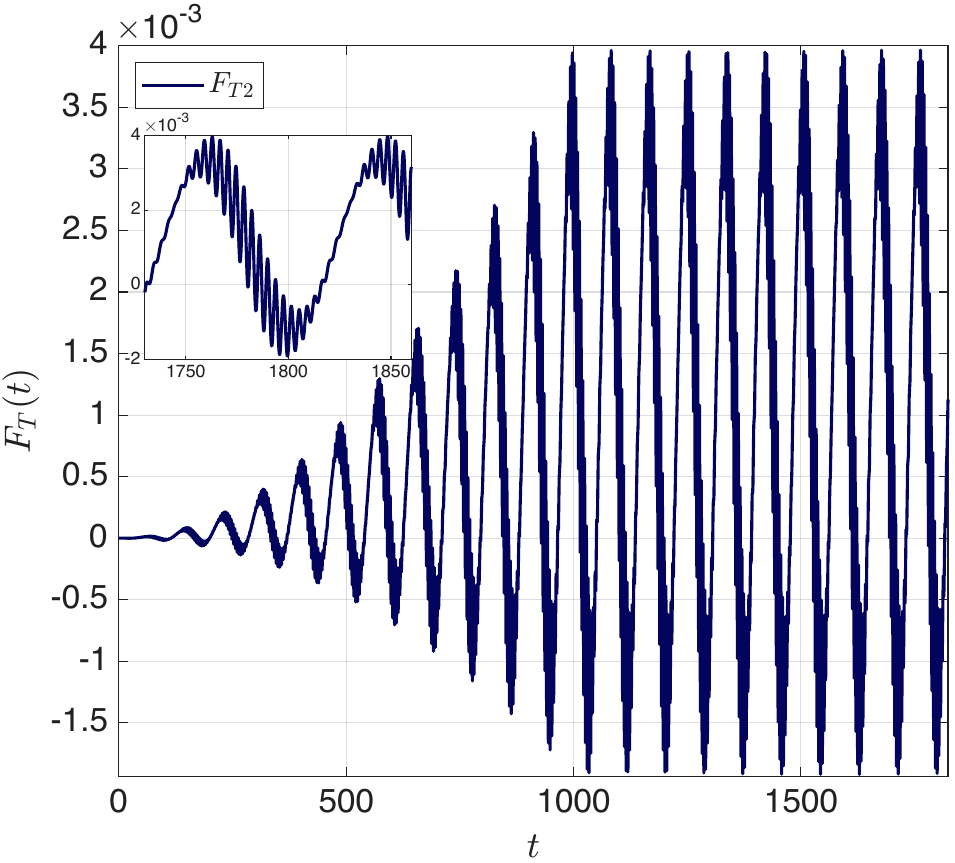}
\caption{\label{fig:SIFt} Force $F_T$ introduced by the phonon to the pinned kink (from Eq. \ref{eq:SIamplROM}). For this plot we used $\epsilon=\epsilon_{min}$, $\alpha=0.495$, $\beta=0.02$.}
\end{figure}
\begin{figure}
\centering
\includegraphics[width=0.6\textwidth]{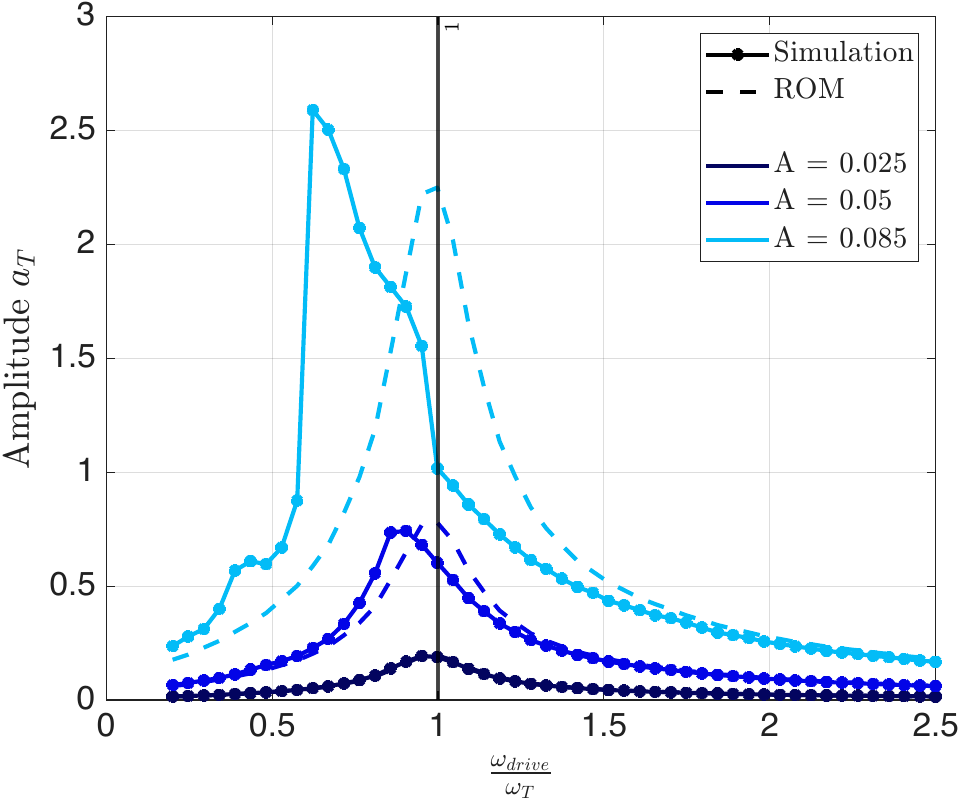}
\caption{\label{fig:SIsoftening} Softening behavior of the oscillating pinned kink. The linear ROM doesn't predict it.  For this plot we used $\epsilon=\epsilon_{min}$, $\alpha=0.495$, $\beta=0.02$.}
\end{figure}

\newpage
\section{Nonlinear Correction to the T Mode ROM}

To account for the position-dependent stiffness observed in the effective pinning potential $E_p(\xi_k)$ from Eq. \ref{eq:pot_en}, where $X_k = \xi_kW$ is the kink position, we extend the reduced-order model for the translational ($T$) mode. This approach captures the asymmetric and varying forcing $\frac{d E_p}{dX_k}$ over large displacements $X(t)$, beyond the validity of local nonlinear corrections.

The mode amplitude $a_T$ is related to the kink position shift $X(t) = X_k - X_p$, where $X_p$ is the static pinned kink position. The translational mode amplitude \(a_T(t)\) is defined as: 
\[a_T(t) = \int_{-\infty}^{\infty} \phi_T\left[ u(x,t) - u_0(x-X_p) \right] dx,\]
where \(u(x,t)\) is the full displacement field and \(u_0(x-X_p)\) is the analytical shape of a static (pinned) kink profile (not the numerically calculated one).  
We now assume the kink moves to a new position \(X(t)\) while retaining its shape \(\big(u(x,t) \approx u_0(x - X(t)-X_p)\big)\). For small position shifts \(X(t)=\Delta X \), we can expand as \(u_0(x - X(t)-X_p) = u_0(x -X_p) + \Delta X \frac{du_0}{dX}\big|_{x - X_p} + \mathcal{O}((\Delta X)^2)\). 
Therefore \(u(x,t) - u_0(x - X_p) \approx X(t) \frac{du_0}{dX}\big|_{x - X_p}\).

Substituting back into the mode amplitude, we can get: \[a_T(t) \approx X(t) \int_{-\infty}^{\infty} \phi_T \frac{du_0}{dX}\big|_{x - X_p} \, dx.\]
The translational mode is proportional to the profile derivative with normalization \(\int_{-\infty}^{\infty} \phi_T^2 \, dx = 1\) and \({m_k} = \int_{-\infty}^{\infty} \left( \frac{du_0}{dx} \right)^2 dx = \frac{1}{3W}\) from the kink's kinetic energy. Therefore \(\phi_T = -\frac{1}{\sqrt{{m_k}}} \frac{du_0}{dX}\big|_{x - X_p}\). 

We can therefore find that: \[a_T(t) \approx -\sqrt{m_k} X(t),\]
with $\frac{d E_p}{da_T}=\frac{d E_p}{dX}\frac{d a_T}{dX}=-\sqrt{m_k}\frac{d E_p}{dX}$. In terms of $a_T$, the model becomes:
\[
\ddot{a_T} + \beta \dot{a_T} + \frac{ \frac{d E_p}{dX}}{\sqrt{m_\text{kink}}} = F_T(t).
\]
For small $a_T$, this reduces to the linear model $\ddot{a_T} + \beta \dot{a_T} + \omega_T^2 a_T = F_T(t)$, with $\omega_T^2 = \frac{d^2 E_p}{dX^2}(0) / m_\text{kink}$.

Although the previous results were obtained for a chain of 250 units, with $\alpha=0.495$, $\beta=0.02$, to focus on the high potential of this method, very similar results can be obtained for other parameters as shown in Fig. \ref{fig:SIoptfreqpred}. When changing the asymmetry $\alpha$ to lower values, the required $\beta$ increases, to maintain the defect strength low. Therefore, higher dissipation and higher $\epsilon$ require higher phonon amplitudes at the boundary to depin the kink. Reducing the length of the systems helps delivering more energy to the pinned kink. We found feasible ($A<(1-\alpha)/2$) to depin the kink for $\alpha=0.45$ and $N=65$. Reducing the length further allows to use higher asymmetries.
\\\\
The optimal excitation frequency can also be approximately found by calculating the period of free oscillation of the kink when it oscillates from a point $X^-$ to $X^+$, where $X^+$ is the position of the saddle point at which depinning happens, and $X^-$ is the other position at the same total energy level as $X^+$. We can define the energy $E_T$ at $X^+$ assuming the system to be conservative (this still approximate the system when damping is low):
\begin{equation}
    E_T=\frac{1}{2}m_k\dot{X^2}+E_P(X).
\end{equation}
We can write the half-period of oscillations as $t_{1/2}=\int_{X^-+\delta}^{X^+-\delta} \frac{dx}{\sqrt{\frac{2}{m_k}[E_T-E_P(X)]}}dx$, where $\delta\ll1$ is a small value to avoid the asymptotic approach to the saddle point. From there the oscillation frequency is $\omega_n=\frac{2\pi}{2t_{1/2}}$. This shows that the softened natural oscillation frequency of the kink approaches $0.5/0.4\omega_T$ in the considered range of $\alpha$.
\\
Interestingly, we predicted and verified that the optimal frequency required remains the same ($\approx 0.4/0.5\omega_T$).
We want to re-iterate that, by using a small defect strength and a short burst of phonon to slow down the kink just before its interaction with the defect (as reported in Fig. \ref{fig:fig3}f of the main text), one can reduce consistently the amplitudes needed to depin the kink, allowing large systems with stronger asymmetries.

\subsection{Sensing Kink Position via Beating Phonons Phase Shift}
As mentioned in the main text, the location of a pinned kink can be sensed by measuring the phase change of the beating phonons from the actuation site to the opposite end of the system. This mechanism leverages the fact that phonons experience a specific phase shift $\delta(\omega)$ when passing through a topological soliton.

Considering two incident phonons introduced at the left boundary ($x=0$) with frequencies $\omega_1, \omega_2$ and amplitudes $a_{P1}, a_{P2}$, the displacement far from the left side of the kink ($x \ll x_d$) is:
\begin{equation}
\delta u_{inc}(x, t) = \mathfrak{Re} [ a_{P1} e^{i(q_{L1}x - \omega_1 t)} + a_{P2} e^{i(q_{L2}x - \omega_2 t)} ],
\end{equation}
where $q_{L1}$ and $q_{L2}$ are the wavenumbers corresponding to the state $u=1$, and the reflected waves are ignored, because the coefficients $R_k\approx 0$ for the range of frequencies, and defect strengths used.

As each phonon propagates through the kink at position $x_d$, its phase shifts $\delta_k(\omega_k)$. For $x > x_d$, the transmitted waves are:
\begin{equation}
\delta u_{trans}(x, t) = \mathfrak{Re} [ a_{P1} T_1 e^{i(q_{R1}x - \omega_1 t + \delta_1)} + a_{P2} T_2 e^{i(q_{R2}x - \omega_2 t + \delta_2)} ],
\end{equation}
where $T_k$ are the transmission coefficients and $q_{Rk}$ are the wavenumbers for the state $u=0$. The total phase of the beating envelope $\Psi(x)$ at any point $x > x_d$ is determined by the difference in the arguments of the two transmitted waves:
\begin{equation}
\Psi(x) = (q_{R1} - q_{R2})x + (\delta_1 - \delta_2).
\end{equation}

By accounting for the travel distance in the left state ($x_d$) and the right state ($L - x_d$), the total phase accumulated at the far end of a domain of length $L$ is:
\begin{equation}
\Psi_{total} = \Delta q_L x_d + \Delta \delta + \Delta q_R (L - x_d),
\end{equation}
where $\Delta q_L = q_{L1} - q_{L2}$, $\Delta q_R = q_{R1} - q_{R2}$, and $\Delta \delta = \delta_1 - \delta_2$. Rearranging this expression, the kink position $x_d$ is analytically determined by:
\begin{equation}
x_d = \frac{\Psi_{total} - \Delta \delta - \Delta q_R L}{\Delta q_L - \Delta q_R}.
\end{equation}

\begin{figure}
\centering
\includegraphics[width=0.6\textwidth]{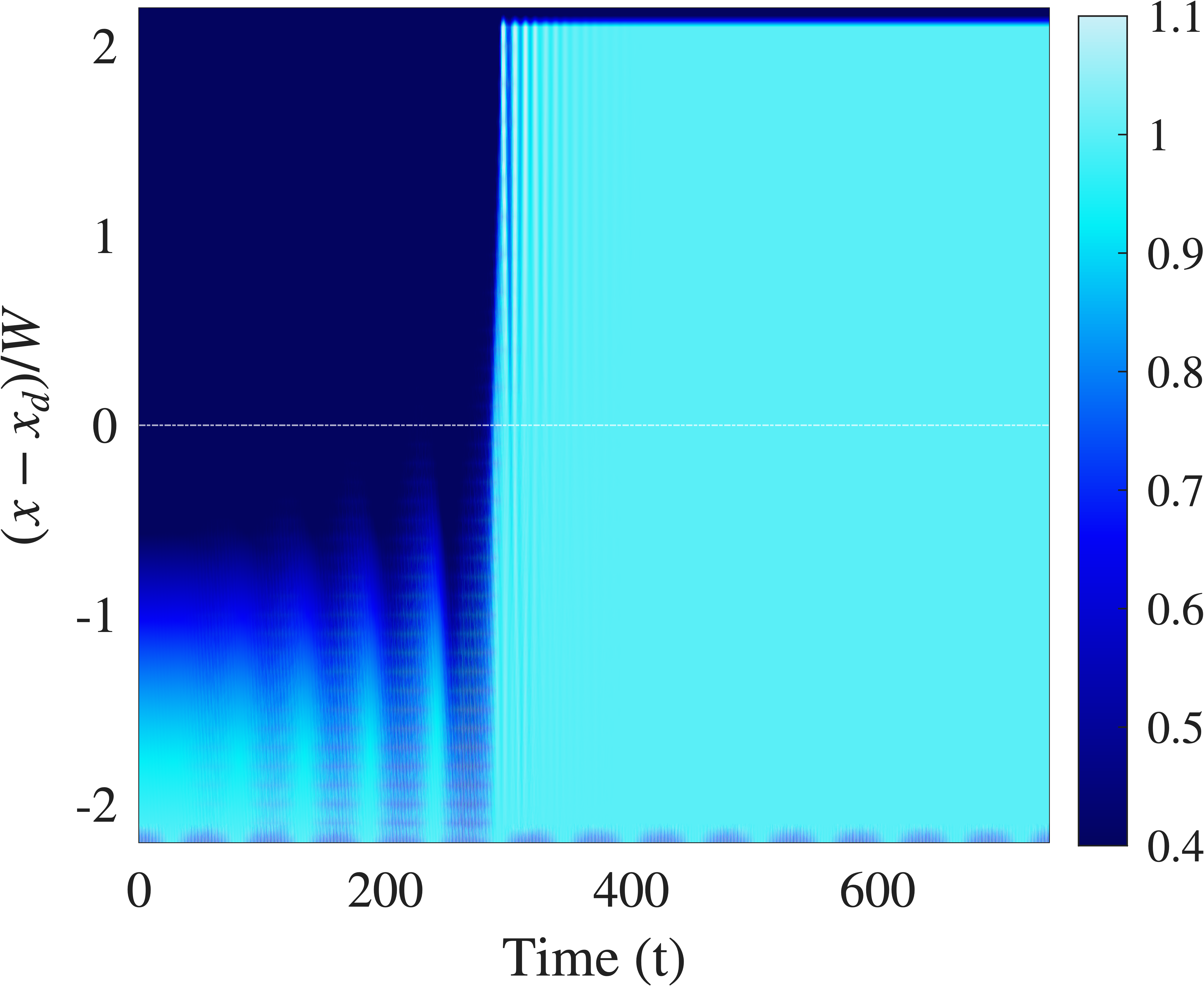}
\caption{\label{fig:SIdepinninga045} Depinning a system with asymmetry $\alpha=0.45$.}
\end{figure}

\begin{figure}
\centering
\includegraphics[width=0.6\textwidth]{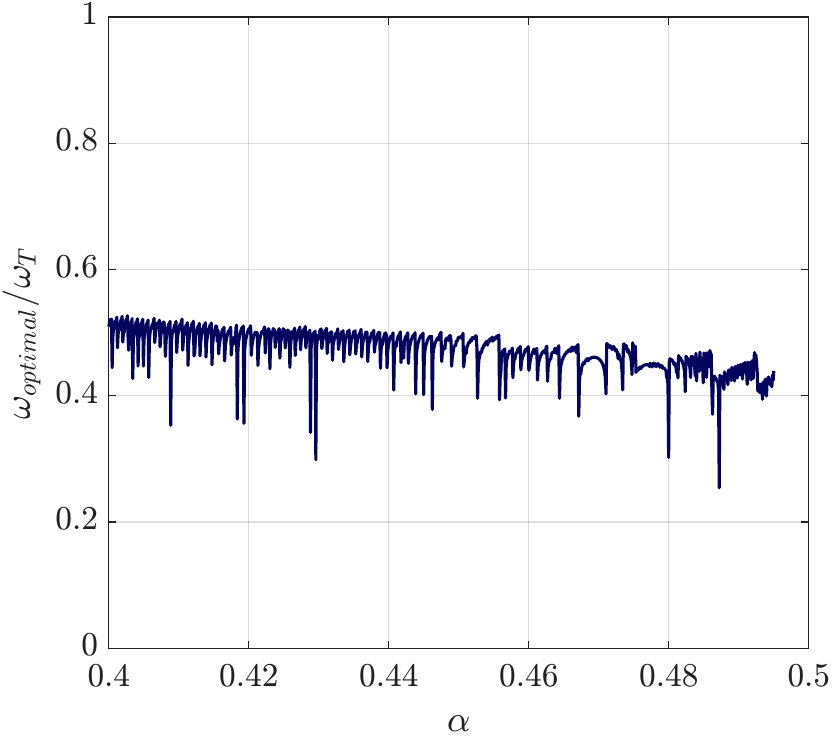}
\caption{\label{fig:SIoptfreqpred} For different $\alpha$, the optimal frequency remains between $0.4\omega_T$ and $0.5\omega_T$.}
\end{figure}

\end{document}